\documentclass[aps,prb,superscriptaddress,showpacs,floatfix,10pt]{revtex4-1}
\usepackage{amsmath}
\usepackage{amsfonts}
\usepackage{graphicx}
\usepackage{epsfig}
\usepackage{psfrag}
\usepackage{color}
\usepackage{epstopdf}
\usepackage{natbib}
\usepackage{xfrac}
\usepackage[normalem]{ulem}
\usepackage{bibentry}
\usepackage{hyperref}
\usepackage{dcolumn}
\usepackage{bm}

\newcommand{\qq}{\mathbf{q}}
\newcommand{\rr}{\mathbf{r}}

\begin{document}

\title{Thermoelectric transport parallel to the planes in a multilayered Mott-Hubbard heterostructure}
\author{Veljko Zlati\'c}
\affiliation{Institute of Physics, Zagreb POB 304, Croatia}
\affiliation{Physics Department, University of Split, Croatia}
\author{J. K. Freericks}
\affiliation{Department of Physics, Georgetown University, 37th and O Sts. NW, Washington, DC 20057, USA}

\begin{abstract}
We present a theory for charge and heat transport parallel to the interfaces of a multilayer (ML) of the ABA type, 
where A and B are materials with strongly correlated electrons. 
When separated, both materials are half-filled Mott-Hubbard insulators with large gaps in their excitation spectrum.  
In a ML,  the renormalization of the energy bands gives rise to a charge reconstruction which breaks the charge neutrality of the planes next to the interface. 
The ensuing electrical field couples self-consistently to the itinerant electrons, so that the properties of the ML crucially depend on an interplay 
between the on-site Coulomb forces and the long range electrostatic forces. 
Using the Falicov-Kimball model, we compute the Green's function and the local charge on each plane of the ML by inhomogeneous DMFT and find 
the corresponding  electrical potential from Poisson's equation. The self-consistent solution is obtained by an iterative procedure,  
which yields the reconstructed charge profile, the electrical potential, the planar density of states, the transport function, and the transport coefficients 
of the device. For the right choice of parameters, we find that a heterostructure built of two Mott-Hubbard insulators exhibits, in a large temperature interval, 
a linear conductivity and a large temperature-independent thermopower. The charge and energy currents are confined to the central part of the ML. 
Our results indicate that correlated multilayers have the potential for applications; by tuning the band shift and the Coulomb correlation on the central planes, 
we can bring the chemical potential  in the immediate proximity of the Mott-Hubbard gap edge and optimize  the transport properties of the device. 
In such a heterostructure, a small gate voltage can easily induce a MI transition. 
Furthermore, the right combination of strongly correlated materials with small ZT can produce, theoretically at least, a heterostructure with a large ZT. 

\end{abstract}


\maketitle

{\section{Introduction}\label{introduction}}

Heterogeneous materials in which a large difference of the chemical potentials gives rise to the emergence of metallic sheets at the interfaces 
have recently attracted a great deal of attention\cite{millis_okamoto}. 
The heterogeneous materials which are insulating  but close to the metal-insulator (MI) transition might be of considerable technological  
interest but have not yet been theoretically studied. 
In this paper, we provide a description of the charge and energy transport in a multilayer (ML)  consisting 
of three stacks of insulating planes. Two identical ones (material A) make up the left and the right semi-infinite leads, 
which are connected to the central section, composed of a different finite stack of insulating planes (material B).  
The leftmost and the rightmost plane of the self-consistently determined ML are attached to semi-infinite bulk materials which set the overall chemical potential of the device.
When disconnected, materials A and B are half-filled Mott insulators with their chemical potentials located in the middle of 
their respective Mott-Hubbard gaps. In the ML, the properties change dramatically and, for the right choice of parameters, 
non-vanishing charge and heat currents emerge parallel to the interface.  

The unusual properties of  such multilayers result from the competition between the short-range Coulomb repulsion and 
long-range electrostatic energy, caused by the electronic charge redistribution. 
To study the ML, we use the spinless Falicov-Kimball model with large electron-electron interaction\cite{falicov_kimball_1969} 
and compute  the Green's functions and the renormalized charge distribution with inhomogeneous dynamical mean-field theory (DMFT). 
The generalization to include spin is straightforward. 
The charge redistribution due to the interfaces, gives rise to a long range electrostatic potential that couples to electrons 
on each plane and affects their dynamics. The self-consistency of the solution is ensured by treating the DMFT equations for 
the Green's functions and the charge density together with Poisson's equation for the electrical potential. 

The isolated subsystems A and B have an excitation spectrum that is symmetric  around their respective chemical potentials 
but the bands in material B are shifted with respect to those in A by an energy $\Delta E$. 
In the ML with a common chemical potential, the local density of states (DOS) on the planes in the leads is nearly the same as in the bulk, 
with the exception of the planes near the interface. On these planes there is an accumulation of charge and, consequently, the local DOS 
deviates from the bulk shape.  In the central part, the local charge on most planes is also the same as in the bulk 
but the symmetry of excitations with respect to the common chemical potential is lost: 
the local DOS is shifted almost rigidly by $-\Delta E$, bringing one of the Hubbard band edges closer to the chemical potential. 
The charge on a few planes next to the interface is reduced and a local DOS on these planes is not just shifted but also distorted.   
Thus, the interfacing gives rise to a screened-dipole layer which greatly reduces the gap in the transport DOS and transforms the ML into 
a small-gap semiconductor with the chemical potential in the proximity of the band edge. An electric field applied perpendicularly to the planes 
can switch the device between semiconducting and metallic states. This switching does not involve the diffusion of electrons over macroscopic 
distances, so that the characteristic time-scale can be much shorter than in usual semiconductor devices. 

The transport properties of the ML are obtained by linear response theory and the thermoelectric response of the ML is calculated  by considering 
the electronic degrees of freedom only. 
Of course, in a ML with semi-infinite leads, the phonon conductivity parallel  to the planes might not be small far enough from  the interface. 
But since the main effect of the mismatch of the chemical potentials due to $\Delta E$ is the shift of the B-bands with respect to the A-bands, 
we expect similar features in a device with finite leads or in an ABA$\cdots$ABA$\cdots$ABA  heterostructure. 

The paper is organized as follows. In Sec.\ref{model}, we introduce the model Hamiltonian of an inhomogeneous multilayer.  
In Sec. \ref{calculations}, we show how to obtain the planar Green's functions by inhomogeneous DMFT and find the self-consistent solution 
that satisfies Poisson's equation. In that section, we also compute the stationary currents by linear response theory, find  the transport function, 
and obtain the transport integrals by the Jonson-Mahan theorem. 
In Sec. \ref{numerical results}, we  present the numerical results for the charge redistribution, the renormalization of the electrical potential,  
 the single-particle spectral function, and show that the gap in the excitation spectrum is much reduced by interfacing.  
We also discuss the results for the transport coefficients and the figure of merit of the device. Sec.\ref{summary} provides the conclusions and summary. \\

{\section{The model Hamiltonian of a correlated multilayer}\label{model}}
We perform self-consistent calculations for a  ML of the ABA type with $N=2L + M$ planes perpendicular to the $z$-axis. 
There are $L$ identical planes (material A) in the left and right part, and $M$ planes (material B) in the central part,  with $M$ assumed to be odd. 
The correlation effects are described by the spinless Falicov-Kimball model with the bulk concentration 
of 1/2  conduction and 1/2 localized electrons per site, and with a large on-site Coulomb repulsion between them. 
The electronic charge on each site is compensated by the background charge of the ionic cores $n_{BG}$ which ensures charge neutrality. 
In the ML,  the translational symmetry is preserved in the $x$ and $y$ directions but it is broken in the $z$ direction. 

The leftmost self-consistent plane of the ML is  labeled by index $\alpha=0$  and the rightmost  one by $\alpha=2L+M-1$. 
These planes are connected to bulk reservoirs which set the overall chemical potential of the ML. 
The plane next to the interface in the left lead, the plane next to the interface in the central part, 
and  the mirror symmetry plane of the ML, are indexed by $\alpha^{-}=L-1$,  $\alpha^{+}=L$, and $\alpha_c=L+(M-1)/2$, 
respectively (with the symmetry mirror in the center of the device).  
In what follows, the planes are denoted by Greek labels and the 2-dimensional vectors parallel to the planes are denoted by Roman labels,  
e.g., $( \alpha,{\bf r} )$  denotes the point on plane $\alpha$ at site $\rr$.

The Hamiltonian of the ML reads\cite{freericks.06,freericks_2007}
\begin{equation}
\mathcal{H}=\mathcal{H}_{\rm 0}+\mathcal{H}_{\rm int}-\mu\mathcal{N}
~, 
				\label{eq: hamtotal}
\end{equation}
where $\mathcal{H}_0$ describes the single-particle Hamiltonian, 
$\mathcal{H}_{\rm int}$ describes the on-site Falicov-Kimball interaction,  
$\mathcal{N}$ is the number operator for conduction electrons, 
and $\mu$  is the chemical potential which determines the total number of conduction electrons in the ML   

The single-particle Hamiltonian has four terms, 
\begin{equation}
\mathcal{H}_0=\mathcal{H}_{\rm T}
+\mathcal{H}_{\rm offset}
+\mathcal{H}_{\rm charge}
+\mathcal{H}_{\rm f}
~.
                         \label{eq: ham_0_total}
\end{equation} 
The first one gives the kinetic energy due to the conduction electrons hopping between neighbouring sites,  
\begin{eqnarray}   
\mathcal{H}_T 
=\sum_{ \alpha,{\bf r}} { h}_T^{ \alpha,{\bf r} }   
~, 
\label{eq: kinetic_symmetric}
\end{eqnarray} 
where the kinetic energy density operator can be written in the symmetrized form as 
\begin{eqnarray}  
{ h}_T^{ \alpha,{\bf r}}
=
-\frac{t^\parallel}{2} 
\sum_{\bf d}
(
c^\dagger_{ \alpha,{\bf r}}c^{}_{{ \alpha,{\bf r}}+{\bf d}}
+
c^\dagger_{{ \alpha,{\bf r}}+{\bf d}}c^{}_{ \alpha,{\bf r}}) 
-
\frac{t^\perp}{2}
\sum_{\delta}
(
c^\dagger_{ \alpha,{\bf r}}c^{}_{{\alpha+\delta,{\bf r}}}
+
c^\dagger_{{ \alpha+\delta,{\bf r}}}c^{}_{ \alpha,{\bf r}}) 
~. 
						\label{eq: r_kinetic_symmetric}
\end{eqnarray}
The summation over ${\bf d}$ and $\delta$ is over the nearest neighbours in plane $\alpha$ 
and the nearest neighbours perpendicular to that plane, respectively. 
For simplicity, the in-plane and the out-of-plane hopping, $t^\parallel$ and $t^\perp$, are 
set to the same value, $t^\parallel=t^\perp=t$, which defines our energy scale.

The offset term describes the shift of the band-centers of the central planes with respect to the leads, 
\begin{equation}
\mathcal{H}_{\rm offset}=-\sum_\alpha\sum_{{\bf r}\in \rm plane}
\Delta E_{\alpha}\ c^\dagger_{\alpha {\bf r}}c^{}_{\alpha {\bf r}}
\label{eq: hoffset}
~, 
\end{equation}
and we take  $\Delta E_{\alpha}=\Delta E$ for the central B layers,  $L\leq \alpha < L+M$, and $\Delta E_{\alpha}=0$ otherwise. 
If the hopping  across the interface is switched off, the Hamiltonian describes two disconnected bulk materials of A and B type 
with a common zero of energy, fixed by the material A. The bands of bulk B are shifted by $\Delta E$ with respect to bulk A and 
their chemical potentials also differ by the same amount. Both materials are assumed to be half-filled Mott insulators.  
In a ML with a unique chemical potential, the mismatch of the electron bands due to $\mathcal{H}_{\rm offset}$ 
gives rise to the deviation of the electronic charge at each site from the bulk value, so that the background charge $n_{BG}$ typically 
doesn't  compensate the renormalized electron charge on individual planes 
and a fully self-consistent charge density needs to be found that restores overall charge neutrality. 
In what follows, the overall chemical potential of the ML is set to zero and used as the origin of the energy axis.  
Energy, like temperature, is measured in units of $t$. 

The uncompensated charges on a given plane, $\delta n_\alpha=n_\alpha-n_{BG}$, give rise to an electrical field 
which affects the electron dynamics on other planes. 
By Gauss law,  a uniform surface charge density on plane $\alpha$ gives rise to a constant electric field perpendicular to that plane,  
$E_\alpha= \delta n_\alpha/\epsilon_\alpha$, where  $\epsilon_\alpha=\epsilon_0 \epsilon_r^\alpha$ is the permittivity of a 
dielectric surrounding plane $\alpha$ ($\epsilon_0$ is the vacuum permittivity and $\epsilon_r^\alpha$ is the relative permittivity).
In a uniform ML, assuming Coulomb gauge, the electrical potential and potential energy on plane $\beta$ corresponding to the field $E_\alpha$ 
are  $ V_\beta^c(\alpha)= \delta n_\alpha  a(\beta-\alpha)/\epsilon_\alpha$  and $V_\beta(\alpha)=e V_\beta^c $, respectively. 
Writing $\delta n_\alpha = e\ \delta \bar\rho_\alpha $, where $e$ is electron charge and $\delta \bar\rho_\alpha$ is the surface electron-number density, 
yields\cite{freericks_2007}  $ V_\beta(\alpha)=\epsilon^\alpha_{Schott}  \delta \bar\rho_\alpha (\beta-\alpha)$, where $\epsilon^\alpha_{Schott}=a e^2/\epsilon_\alpha$. 
This long-range potential energy has to be taken into account when discussing electron dynamics on plane $\beta$.  
In a ML of the ABA type, the dielectric constant changes between some planes which makes the expression for $ V_\beta(\alpha)$ somewhat 
more complicated\cite{freericks_2007} but the reasoning is the same: the electric potential on plane $\alpha$ and the charge density 
on all other planes have to satisfy Poisson's's equation 
\begin{equation}
\frac{d^2 V^c_\alpha}{d z^2}=-\sum_{\beta\neq \alpha} \frac{\delta n_\beta }{\epsilon_\beta} \delta({z}_\alpha-{z}_\beta )~, 
\label{eq: Poisson}
\end{equation}
where the differential operator and the delta function are defined on a discrete one-dimensional space. 
Since the electron dynamics on every plane depends on the 
charge distribution on all other planes, the quantum mechanical problem, posed by the Hamiltonian, 
and the electrostatic problem, posed by Poisson's equation, have to be solved in a self consistent way. 
The additional potential energy due to the redistribution of charges on the ML planes  
shifts the local electro-chemical potential of itinerant  electrons on every plane 
and contributes the following term to the Hamiltonian  
\begin{equation}
\mathcal{H}_{\rm charge}=\sum_\alpha V_\alpha \sum_{{\bf r}\in {\rm plane}}
c^\dagger_{\alpha {\bf r}}c^{}_{\alpha {\bf r}}
~.
\label{eq: charge_ham} 
\end{equation}
In our calculations, the number of planes in the leads has to be large enough that effect of the electronic charge reconstruction 
on the first few and the last few planes of the ML can be neglected. 

The localized $f$-electrons are represented by a set of energy levels and they are described by the Hamiltonian 
\begin{equation}
\mathcal{H}_{\rm f}=\sum_\alpha \sum_{{\bf r}\in {\rm plane}}
E_{\alpha {\bf r}} f^\dagger_{\alpha {\bf r}}f^{}_{\alpha {\bf r}} ~, 
\label{eq: f_ham}
\end{equation}
where we assume that the distribution of the $f$-electrons over the available energy levels is random 
and that the translational symmetry is restored by  averaging $n_\alpha^f$ over all possible configurations. 
The concentration of $f$-electrons is permanently fixed at half filling.
Finally, the interaction between the conduction and localized electrons is described by the term
\begin{equation}
\mathcal{H}_{\rm int}
=
\sum_\alpha \sum_{{\bf r}\in {\rm plane}}
U_{\alpha {\bf r}} \
c^\dagger_{\alpha {\bf r}}c^{}_{\alpha {\bf r}} \ 
f^\dagger_{\alpha {\bf r}}f^{}_{\alpha {\bf r}}  
~
\label{eq: f_ham}
\end{equation}
where $U_{\alpha {\bf r}} $ is the short range Coulomb interaction at site $(\alpha {\rr})$ and we take $U_{\alpha {\bf r}} =U_l$  in the leads 
(material A) and $U_{\alpha {\bf r}} =U_c$ in the central part (material B).  
Most of the numerical results are obtained for $U_l=U_c=U\gg t $\\

{\section{Calculation of local Green's functions and transport coefficients by inhomogeneous DMFT }\label{calculations} }
Sec.~\ref{Green's functions} summarizes the derivation of the Green's function for an inhomogeneous model described by the Hamiltonian \eqref{eq: hamtotal}. 
In  Sec.~\ref{DMFT}, we introduce the inhomogeneous DMFT to compute the renormalized charge on each plane and use Poisson's equation to find the 
corresponding electrical potential. Redefining $\mathcal{H}_{\rm charge}$ by this potential, we recalculate the Green's function and iterate 
the whole procedure to a fixed point.  
In Sec.~\ref{transport function}, the self-consistent solution for the Greens function is used to calculate the transport function 
and, eventually, the Jonson-Mahan theorem is used to obtain the transport coefficients. \\

{ \subsection{Calculation of the Green's functions}\label{Green's functions}}
The local DOS on each plane and the transport coefficients of the ML are obtained from 
the single-particle Green's functions. 
Since the Hamiltonian is time-independent, the equation of motion (EOM)  
for the Green's function can be written in operator form as 
\begin{equation}
(z-H)G=I
~
\end{equation}
with z a complex variable in the upper half plane. 
For the ML with translational symmetry parallel to the layers, 
the matrix elements of the renormalized Green's function satisfy the following EOM in real space
\begin{eqnarray}
zG^{}_{\alpha\beta}({\bf x}-{\bf y},z) - 
\sum_{\gamma{\bf r}} H_{\alpha\gamma}({\bf x}-{\bf r}) 
G_{\gamma\beta}({\bf r}-{\bf y},z)
=\delta_{\alpha\beta} \delta({\bf x}-{\bf y})~, 
\label{eq: EOM_real_space}
\end{eqnarray}
where the summation is over all the planes $\gamma$ and all the sites ${\bf r}$ in that plane.
Using the translational invariance within the planes, we make the in-plane Fourier transform, 
introduce the mixed representation\cite{potthoff_nolting_1999}, 
and write the matrix elements of the Green's function as  $G_{\alpha\beta}({\bf q},z)
=
\sum_{\bf x} G_{\alpha\beta}({\bf x-y},z) \exp\{i{\bf q}\cdot{(\bf x-y)}\}$, 
where $\bf q$ is a 2-dimensional vector in reciprocal space.
For the noninteracting Green's function, obtained from the Hamiltonian in Eq. \eqref{eq: hamtotal} with $U=0$,  
the EOM  in the mixed representation becomes, 
\begin{eqnarray}
\sum_{\gamma}
\left[
[z+\bar\mu_\alpha-\epsilon^{\parallel}({\bf q})]
\delta_{\alpha\gamma}
+
t \delta_{\alpha-1,\gamma}
+
t \delta_{\alpha+1,\gamma}
\right]
 G^{0}_{\gamma\beta}({\bf q},z)
=\delta_{\alpha\beta} 
~, 
                             \label{eq: EOM_0_hop}
\end{eqnarray}
where  
$\bar\mu_\alpha=\mu+\Delta E_\alpha-V_\alpha$ 
is the local electrochemical potential, 
$\epsilon^{\parallel} ({\bf q})=
\sum_{\bf r} t^{\parallel}  \exp\{i{\bf q}\cdot{\bf r}\}
$ 
is the in-plane dispersion, 
and the square bracket defines the inverse matrix $[{\bf G}^{0}]^{-1}_{\alpha\beta}({\bf q},z)$ with respect to the planar labels. 
Because of translational invariance within the planes and with respect to time,  the Dyson equation for the interacting Green's function 
can be written as,  
\begin{eqnarray}
{\bf G}_{\alpha\beta}({\bf q},z) 
=
{\bf G}_{\alpha\beta}^{0}({\bf q},z)
+
\sum_{\gamma\delta}
{\bf G}_{\alpha\gamma}^{0}({\bf q},z)
{\bf \Sigma}_{\gamma\delta}({\bf q},z) {\bf G}_{\delta\beta} ({\bf q},z) ~, 
                       \label{eq: Dyson}
\end{eqnarray}
where ${\bf \Sigma}_{\alpha\beta}({\bf q},z)$  is the non-local self-energy matrix. 
We now make a crucial approximation of the DMFT and assume that the self energy is local, 
${\bf \Sigma}_{\alpha\beta}({\bf q},z) = {\bf \Sigma}_{\alpha}(z)\delta_{\alpha\beta}$. 
This certainly holds for homogeneous systems in infinite dimensions, where the non-local corrections to the self-energy vanish.
In a 3d ML, the importance of the non-local corrections is not really known but the local approximation allows us to solve the inhomogeneous problem 
and obtain an insight into the properties of the system. 
By multiplying Eq.~(\ref{eq: Dyson}) from the left by the inverse of ${\bf G}_{\gamma\beta}^{0}({\bf q},z)$, summing over $\gamma$,  
and using Eq.~\eqref{eq: EOM_0_hop},  we reduce the EOM for the renormalized Green's function  to a simple form
\begin{eqnarray}
\left[
z+\bar\mu_\alpha-\Sigma_{\alpha}(z) -\epsilon^{\parallel}_{}({\bf q})
\right]
 G^{}_{\alpha\beta}({\bf q},z)
+
t \;G^{}_{\alpha-1\beta}({\bf q},z)
+
t \; G^{}_{\alpha+1\beta}({\bf q},z)
=\delta_{\alpha\beta} ~
                        \label{eq: EOM_hop}
\end{eqnarray}
which represents a discrete 1-dimensional problem that can be solved recursively\cite{economou}.

The diagonal part of the renormalized Green's function is obtained from the above equation for $\alpha=\beta$, 
\begin{eqnarray}
 G^{}_{\alpha\alpha}({\bf q},z)
=
\frac{1}{L_{\alpha}({\bf q},z) + R_{\alpha}({\bf q},z) -
\left[
z+\bar\mu_\alpha-\Sigma_{\alpha}(z) -\epsilon^{\parallel}_{}({\bf q}) 
\right]
}
\equiv G^{}_{\alpha}({\bf q},z)
~, 
                      \label{eq: G_alpha_alpha}
\end{eqnarray}
where auxiliary functions $L_{\alpha-n}$ and $R_{\alpha-n}$ satisfy the recursion relations 
\begin{eqnarray}
 L_{\alpha-n}({\bf q},z)
&=&
z+\bar\mu_\alpha-\Sigma_{\alpha-n}(z) -\epsilon^{\parallel}_{}({\bf q}) -
\frac{t^2 } {L_{\alpha-n-1}({\bf q},z)}
 ~,
\label{eq: L-recursion}
\\ 
R_{\alpha+n}({\bf q},z)
&=&
z+\bar\mu_\alpha-\Sigma_{\alpha+n}(z) -\epsilon^{\parallel}_{}({\bf q}) -
\frac{ t^2} {R_{\alpha+n+1}({\bf q},z)}
 ~.
\label{eq: R-recursion}
\end{eqnarray}
Assuming that far enough from the interface the planar Green's functions of the ML 
are the same as in the bulk, we approximate 
$ L_{\alpha-n}({\bf q},z)\simeq  L_{\alpha-n-1}({\bf q},z) \simeq L_{0}({\bf q},z)$ and 
$ R_{\alpha+n}({\bf q},z)\simeq R_{\alpha+n+1}({\bf q},z)\simeq  R_{N}({\bf q},z)$. 
This reduces Eqs.~(\ref{eq: L-recursion}) and (\ref{eq: R-recursion}) for the leftmost and the rightmost 
auxiliary functions  to quadratic forms  
with the solutions  
\begin{eqnarray}
 L_{0}({\bf q},z)
&=&
\frac{ z+\bar\mu_\alpha-\Sigma_{0}(z) -\epsilon^{\parallel}_{}({\bf q}) }{2} 
\pm 
\sqrt{[z+\bar\mu_\alpha-\Sigma_{0}(z) -\epsilon^{\parallel}_{}({\bf q})]^2-4t^2}
 ~,
\label{eq: L-0}
\\
 R_{N}({\bf q},z)
&=&
\frac{ z+\bar\mu_\alpha-\Sigma_{N}(z) -\epsilon^{\parallel}_{}({\bf q}) }{2} 
\pm 
\sqrt{[z+\bar\mu_\alpha-\Sigma_{N}(z) -\epsilon^{\parallel}_{}({\bf q})]^2-4t^2}
 ~,
\label{eq: R-0}
\end{eqnarray}
where $\Sigma_{0}(z) $ and $\Sigma_{0}(N) $  are the same as the self energy in  bulk A.  
Thus, if we know $\Sigma_{\alpha}(z)$ on each plane, we can generate the auxiliary functions $ L_{1}, L_{2}, \ldots L_{N}$ and  
$ R_{N-1}, R_{N-2}, \ldots R_{0}$ starting from $ L_{0}$ and $ R_{N}$.
For example, $ L_{1}$ and $ R_{N-1}$ are obtained by setting $n=\alpha-1$ in Eq.~(\ref{eq: L-recursion}) and  
$n=N-\alpha-1$ in Eq.~(\ref{eq: R-recursion}).
Knowing $ L_{\alpha}$, $ R_{\alpha}$, and $\Sigma_{\alpha}(z)$ for each plane, 
we get the diagonal Green's function from Eq.~(\ref{eq: G_alpha_alpha}) 
and the off-diagonal one from  Eq.~ (\ref{eq: EOM_hop}). That is 
\begin{eqnarray}
G_{\alpha+1,\alpha}({\bf q},z)
= 
- \frac{L_{\alpha}({\bf q},z)}{t_{}^{} } G_{\alpha,\alpha}({\bf q},z)
 ~
\label{eq: G_alpha+1}
\end{eqnarray}
and 
\begin{eqnarray}
G_{\alpha-1,\alpha}({\bf q},z)
= 
- \frac{R_{\alpha}({\bf q},z)} {t_{}^{} }G_{\alpha,\alpha}({\bf q},z)
 ~.
\label{eq: G_alpha-1}
\end{eqnarray}
Continuing this recursive procedure yields $G_{\alpha\beta}$ for all $\alpha$ and $\beta$.
Since the Green's function and auxiliary functions depend on the planar momentum only via  $\epsilon^ {}({\bf q}) $, 
we can calculate the local Green's function for plane $\alpha$ by replacing the 2-dimensional momentum summations 
by an integral over the corresponding 2-dimensional density of states, 
\begin{eqnarray}
\rho_{2D}(\epsilon)
=
\sum_{\bf q} ~ \delta(\epsilon -\epsilon^{\parallel}(\bf q) ) ~.
\label{eq: rho_2D}
\end{eqnarray}
The diagonal part of the local Green's function for plane $\alpha$ follows from Eq.~(\ref{eq: G_alpha_alpha}), which gives 
\begin{eqnarray}
 G^{}_{\alpha}(z)
=
\int d\epsilon_ {}~\rho_{2D}(\epsilon_ {})
\frac{1}{L_{\alpha}(\epsilon_ {},z) + R_{\alpha}(\epsilon_ {},z) -
\left[
z+\bar\mu_\alpha-\Sigma_{\alpha}(z) -\epsilon^ {} 
\right]
}~,
                      \label{eq: G_alpha}
\end{eqnarray}
and the off-diagonal parts are obtained by integrating Eqs.~\eqref{eq: G_alpha+1} and \eqref{eq: G_alpha-1}. \\

{\subsection{The inhomogeneous DMFT solution}\label{DMFT}}
The DMFT solution for the local Green's functions and the self-energy  of the multilayer are obtained\cite{potthoff_nolting_1999,freericks.06} 
by equating $G^{}_{\alpha}(z)$ and $\Sigma_{\alpha}(z)$ in Eq.~\eqref{eq: G_alpha} with the Green's function and the self-energy 
of a single-site Falicov-Kimball model with the same parameters as for the plane $\alpha$ of the lattice. 
The mapping of the lattice model with $2L+M$  planes on $2L+M$ single-site Falicov-Kimball models allows us to find the solution 
 by an iterative procedure. This solution is exact for a homogeneous system in infinite dimensions, where the self-energy functionals 
of the lattice and the single-site model are defined by identical momentum-independent skeleton diagrams. 
In a 3d ML, the mapping holds to the extent that the momentum-dependence of all the self-energy diagrams can be neglected. 

To obtain the DMFT solution, we make an initial guess for $ \Sigma_{\alpha}(z) $ on each plane of the ML,  
generate the auxiliary functions $L_{\alpha}$ and $ R_{\alpha}$, and calculate $G_{\alpha}(z)$ using Eq.~\eqref{eq: G_alpha}. 
Then, using the Dyson equation, we define the inverse of the unperturbed single-site Green's function associated with plane $\alpha$ as, 
\begin{eqnarray}
 [G_{0\alpha}^{imp}(z)]^{-1}
=
 [G_{\alpha}(z)]^{-1} + \Sigma_{\alpha}(z) ~. 
                          \label{eq: G_0_alpha}
\end{eqnarray}
The inverse of Eq.~(\ref{eq: G_0_alpha}) yields $G_{0\alpha}^{imp}(z)$  which we substitute into the exact expression for 
the renormalized Green's function of the single-site Falicov-Kimball model
\begin{eqnarray}
 G_{\alpha}(z)
=
\frac{G_{0\alpha}^{imp}(z)}{2} +
\frac{1/2}{ [G_{0\alpha}^{imp}(z)]^{-1} - U_\alpha } 
                       \label{eq: G_FK_alpha} ~ .
\end{eqnarray}
The fact that the functional form of the single-site Green's function is known exactly makes the Falicov-Kimball lattice much easier 
to solve than the Hubbard or Anderson lattices. In the latter case, the impurity Green's function can only be found numerically,  
which renders the multilayer problem challenging. 
Using $G_{0\alpha}^{imp}(z)$ and $G_{\alpha}(z)$, we recalculate the impurity  self-energy 
 \begin{eqnarray}
  \Sigma_{\alpha}(z) =  [G_{0\alpha}^{imp}(z)]^{-1} - [G_{\alpha}(z)]^{-1} 
 ~ , 
                          \label{eq: sigma_new_alpha}
\end{eqnarray} 
substitute it back into Eq.~\eqref{eq: G_alpha} for the local Green's function, and iterate this procedure to a fixed point. 
The converged result yields the self-energy, the auxiliary functions  $L_{\alpha}(\epsilon^ {},z)$ and $ R_{\alpha}(\epsilon^ {},z)$, 
and  the diagonal and off-diagonal Green's functions on all the planes.   

The diagonal Green's function, obtained by the DMFT for a given choice of electrostatic potentials (given band offsets) 
in Eq.\eqref{eq: charge_ham}, provides the local charge on plane $\alpha$,  
\begin{equation}
n_\alpha=\int d\omega~f(\omega)\  \rho_\alpha(\omega) 
                       \label{eq: n_alpha} ~ ,
\end{equation}
where $f(\omega)=1/(\exp(\omega/T)+1)$ is the Fermi function and $ \rho_\alpha(\omega)= - {\rm Im} ~G_\alpha(\omega) /{\pi}$ is the local DOS on plane $\alpha$.  
Using this renormalized charge in Poisson's equation \eqref{eq: Poisson}, we  compute the renormalized potential and  
substitute it into the offset term of the Hamiltonian for the next cycle of calculations.  
The fixed point of the full iterative procedure (DMFT+Poisson) yields the self-consistent solution of the coupled quantum mechanical and electrostatic problems. 
The $f$-electron charge is unrenormalized by these calculations, i.e. we always have $n^f=1/2$. 

In a homogeneous system with a uniform charge distribution, where Poisson's equation is automatically satisfied, 
the iterative solution of DMFT equations approaches rapidly the fixed point;  the local Green's function $G_{bulk}(\omega)$ is obtained in a few iterations. 
In a heterostructure, where  the on-site Coulomb potential competes with the long-range electrostatic one, 
the local electro-chemical potential and the charge on each plane have to by adjusted self-consistently, which slows down the convergence. 
The fixed point is now approached very slowly, if ever. (See discussion at Sec.\ref{charge reconstruction}.) 
Furthermore, we need a large number of planes in the leads, so  that the renormalization of the electrical potential can be neglected 
at the first and the last plane of the ML, and  $G_0$ and $G_{N-1}$ can be approximated by $G_{bulk}$. 
All this makes the numerical calculations for the heterostructure quite demanding.
Once we find the self-consistent values of $n_\alpha$ and $V_\alpha$ by solving the DMFT and Poisson's equations on the imaginary axis, 
we can calculate the self-energy on the real axis by iterating DMFT equations Eqs.~\eqref{eq: G_alpha} -- \eqref{eq: sigma_new_alpha} 
without Poisson's loop.\\

{\subsection{Linear response theory and the transport function}\label{transport function}}
The transport properties of a ML are obtained by computing the macroscopic 
currents with linear response theory. The charge current density in the planes perpendicular to the z-direction is 
\begin{equation}
                                  \label{eq:qm_j}
{\bf J}({\bf x}) = {\mathrm Tr} \{ \rho_\phi \hat{\bf j}_0({\bf x})\},
\end{equation}
where $\rho_\phi $ is the density matrix of the particles in an external electrical field  $ {\bf E}_\phi$ parallel to the planes 
and  $ \hat{\bf j}_0({\bf x})$ is the current density operator\cite{mahan.81}. 
An expansion of $\rho_\phi $ to lowest order in $ {\bf E}_\phi$ yields 
\begin{eqnarray}  \hskip -0.5cm
                               \label{current-x}
{\bf J}({\bf x})
&=&
e^{i\omega t}
\int d{\bf x}^\prime ~ {\bf E}_\phi({\bf x}^\prime) 
\int_0^{\infty} dt^\prime e^{-i\omega t^\prime}  
\int_0^\beta d\beta^\prime 
\langle \; \hat {\bf j}({\bf x}^\prime,-t^\prime-i\beta^\prime)  
\hat{\bf j}({\bf x})\rangle_0 ~, 
\end{eqnarray}  
where  $\langle\cdots \rangle_0$ denotes the thermodynamic average with respect to the density matrix $\hat\rho_0$ 
defined in the absence of the external field. 
The static conductivity is obtained by Fourier transforming Eq.~(\ref{current-x}) and taking the ${\bf q}\to 0$ limit before $\omega\to 0$ limit, 
which gives~\cite{luttinger.64} 
\begin{equation} 
                              \label{Kubo-q=0}
\sigma^{}
=V \lim_{\omega\to 0}
\int_0^{\infty} dt^\prime e^{-i\omega t^\prime} 
\int_0^\beta d\beta^\prime 
\langle \; \hat {\bf j}_{0}^{}(-t^\prime-i\beta^\prime) 
\hat {\bf j}_{0}^{}\rangle_0 ~
\end{equation} 
where $\hat {\bf j}_{0}$ is the uniform ($\qq=0$) component of the current density operator,  
\begin{eqnarray}   
                          \label{eq: j_0}
\hat {\bf j}_{0}=\sum_{{\bf q }\alpha}
{\bf v}_{\bf q }
c^\dagger_{{\bf q}\alpha}c_{{\bf q}\alpha} ~, 
\end{eqnarray}
and ${\bf v}_{\bf q }=\nabla_{\bf q} \epsilon^{\parallel}({\bf q})$ is the in-plane band velocity. 
Note, the electrical potential caused by the charge reconstruction does not drive any current. This is obvious on  physical ground but 
it also follows\cite{freericks_2007} from the exact result for the current density operator defined by the Hamiltonian \eqref{eq: hamtotal}.

We compute the current-current correlation function in Eq.~\eqref{Kubo-q=0} by neglecting the vertex corrections and obtain 
for the conductivity parallel to the ML planes the result 
$$\sigma^{}=\sum_{\alpha\beta} \sigma_{\alpha\beta} ~,$$
where $ \sigma_{\alpha\beta}$ is the off-diagonal conductivity matrix given by 
\begin{eqnarray} \hskip -.5cm
\sigma_{\alpha\beta}
 &=&
\frac{e^2}{2} \sum_{{\bf q}} { {\bf v}_{\bf q } }^2
\int_{-\infty}^{\infty}d\omega~
\left(-\frac{\partial f(\omega)}{\partial \omega}\right)~
[\rm Im ~ G_{\alpha\beta}({\bf q})(\omega)]^2~. 
\label{eq: l11_real}
\end{eqnarray} 
Since the off-diagonal Green's function $ G_{\alpha\beta}({\bf q})(\omega)$ depends on the planar momentum only through the planar energy $\epsilon^\parallel({\bf q})$, 
the ${\bf q}$-summation can be performed by using the 2-D transport DOS, 
\begin{eqnarray}
\rho^{2D}_{tr}(\epsilon)
=
\sum_{\bf q} ~{\bf v_{\bf q} }^2 \delta\left(\epsilon -\epsilon^\parallel({\bf q})\right) ~. 
\label{eq: rho_tr}
\end{eqnarray}  
For the square lattice, the transport DOS is obtained by solving the equation 
$$
\frac{d \rho^{2D}_{tr}(\epsilon)}{d\epsilon}=-\frac{\epsilon}{4} \; \rho_{2D}(\epsilon), 
$$
with the boundary condition $\rho^{2D}_{tr}(\epsilon)=0$ at the bottom of the conduction band.

Next, we introduce the transport function for plane $\alpha$, 
\begin{eqnarray}
\Lambda^{\alpha}_{tr}(\omega)
=
\sum_{\beta}
\int d\epsilon ~
\rho^{2D}_{tr}(\epsilon)
[\rm Im ~ G^{}_{\alpha\beta}(\epsilon,\omega)]^2~, 
                      \label{eq: lambda_tr_alpha}
\end{eqnarray}
and write the conductivity of plane $\alpha$ as, 
\begin{eqnarray}
\sigma_{\alpha} 
=
e^2
\int_{-\infty}^{\infty}d\omega~
\left(-\frac{\partial f(\omega)}{\partial \omega}\right)
\Lambda^{\alpha}_{tr}(\omega)
~. 
                           \label{eq: cond_static}                             
\end{eqnarray}
In a ML,  $\Lambda^{\alpha}_{tr}$ depends on the off-diagonal matrix elements of the Green's function, i.e.,  
the current on plane $\alpha$ has contributions due to the excursion of conduction electrons to all other planes. 
Thus, the total static conductivity for transport parallel to the multilayer planes  becomes 
\begin{eqnarray}
\sigma(T)
=
\sum_{\alpha} 
\sigma_{\alpha}
=
e^2
\int_{-\infty}^{\infty}d\omega~
\left(-\frac{\partial f(\omega)}{\partial \omega}\right)
\Lambda_{tr}(\omega)
~,
                      \label{eq: sigma}
\end{eqnarray}
where we introduced the transport function of the ML, 
\begin{eqnarray}
 \Lambda^{}_{tr}(\omega)
 = 
 \sum_\alpha\Lambda^{\alpha}_{tr}(\omega)
 =
 \sum_{\alpha\beta}
\int d\epsilon ~
\rho^{2D}_{tr}(\epsilon)
[\rm Im ~ G^{}_{\alpha\beta}(\epsilon,\omega)]^2~.
                       \label{eq: lambda_tr}
\end{eqnarray}

The Falicov-Kimball model satisfies the Jonson-Mahan theorem,\cite{jonson_mahan_1990,freericks_2007} 
so that all the transport integrals follow immediately from the transport function calculated above. 
We have 
\begin{eqnarray}
L_{mn}(T)
=
\int_{-\infty}^{\infty}d\omega~
\left(-\frac{\partial f(\omega)}{\partial \omega}\right)\ \omega^{m+n-2} \ 
\Lambda^{}_{tr}(\omega)~ 
                           \label{eq: transport_integrals}                             
\end{eqnarray}
which gives the electrical conductivity 
\begin{eqnarray}
\sigma(T)
=
e^2 L_{11}
~,
                      \label{eq: sigma}
\end{eqnarray}
the thermal conductivity
\begin{eqnarray}
\kappa_e(T)
=
\left(\frac{k_B}{e}\right)^2\frac{\sigma(T)}{T}
\left(
\frac{L_{22}}{L_{11}} - \frac{L^2_{12}}{L^2_{11}}
\right)~
                           \label{eq: thermal_conductivity}                             
\end{eqnarray} 
and the Seebeck coefficient  
\begin{eqnarray}
S(T)
=
\left(\frac{k_B}{e}\right)
\frac{L_{12}}{T L_{11}}~.
                           \label{eq: thermopower}                
\end{eqnarray}
The Lorenz number is ${\cal L}=\kappa_e/\sigma T$. 
The efficiency of a particular thermoelectric material depends on the 
dimensionless figure-of-merit, $ZT=S^2\sigma T/\kappa$, 
where  $\kappa=\kappa_e+\kappa_{ph}$ is the overall thermal conductivity 
due to the electronic and  the lattice degrees of freedom.  
The efficient thermoelectric conversion requires $ZT > 1$. 
Neglecting $\kappa_{ph}$, we write the upper bound for the figure-of-merit in terms of the transport integrals as\cite{zlatic_2014b}
\begin{equation}
                                       \label{ZT_numerical}
ZT
=
\frac{L_{12}^2}{L_{22}L_{11}-L_{12}^2} = \frac{S^2}{\cal L}
 ~.
\end {equation}
\\

{\section{Numerical results}\label{numerical results}} 
We now discuss typical features of the electronic charge reconstruction, the renormalization of the electro-chemical potential, 
the local DOS, the transport DOS, and the transport coefficients obtained from the self-consistent solution of the DMFT and Poisson's equation.  
For a ML of the ABA type, the calculations are numerically demanding, because a large number of A planes is needed to ensure 
that the potential decays from its maximum (or minimum) at $\alpha_c$  to $V_{bulk}\simeq 0$ at $\alpha=0$, which is a necessary condition for our solution to be valid.
In order to establish the general trends, we studied the ML with the number planes varying from L=31 to L=51 in the leads and M=5 to M=31 in the center,  
for several band-offsets in the central planes, several values of $U$  on the leads and the center, and various temperatures.  
Most of  the data shown below are obtained for $U=8$t,  $\Delta E=0$  in the leads and  $U=8$t, $\Delta E=-0.9$t in the center. 
The parameter $\epsilon_{Schott}$ controls how fast the electric field decays away from a uniformly charged plane; 
to describe the dielectrics with relative permittivity between 10 and 100 and the lattice spacing of 2 and 5 $a_B$, we 
take $\epsilon_{Schott}=0.4$ in the leads and vary $\epsilon_{Schott}$ from $0.4$ to $50$ in the center. 
(For details regarding the effect of the screening length 
$\epsilon_{Schott}$  on the charge reconstruction see Ref. \onlinecite{freericks.06} and the discussion following Fig. \ref{fig:charge_eschott}.) 
Homogeneous bulk materials with the same parameters as in our calculations would be half-filled Mott insulators. 
As long as the number of A planes is such that the inhomogeneity at the interface does not affect appreciably the leftmost 
($\alpha=0$) and the rightmost ($\alpha=N-1$) plane of the ML, the qualitative features turn out to be insensitive to the number of planes in the leads;  
they are also robust with respect to the number of planes in the center. 

We studied most extensively the ML with $L=41$ and $M=11$, so that the A plane next to the interface is at $\alpha^-=40$, 
the first B plane is at $\alpha^+=41$, and the mirror-symmetric plane  is at $\alpha_c=46$. 
When the leads are disconnected from the central part, all the planes are electrically neutral and we have $n_\alpha=1/2$ on every site 
(assuming a periodic boundary condition for each part). 
In the ML,  the local DOS on the first and the last plane satisfy $\rho_{0}(\omega)=\rho_{N-1}(\omega)=\rho_{bulk}(\omega)$, 
where $\rho_{bulk}(\omega)$ is the local DOS of a uniform system with the same  parameters as in the leads. Since $\Delta E=0$ in the leads, 
the  corresponding bulk is at half-filling and $\rho_{0}(\omega)$ is electron-hole symmetric with respect to the chemical potential. 
\\

\begin{figure*}[thb!]
\centering
\includegraphics[width=0.40\columnwidth,clip]{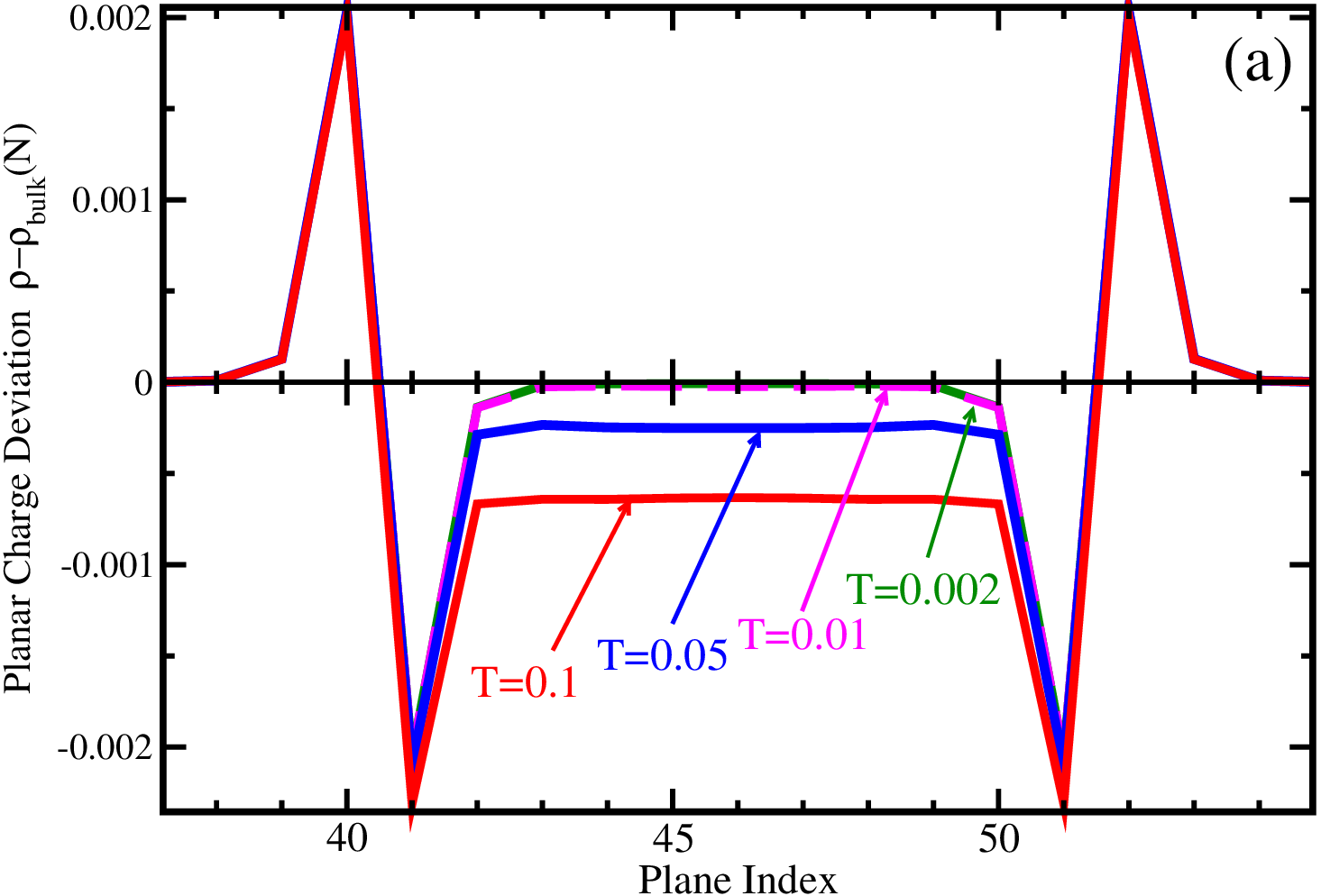}
\includegraphics[width=0.40\columnwidth,clip]{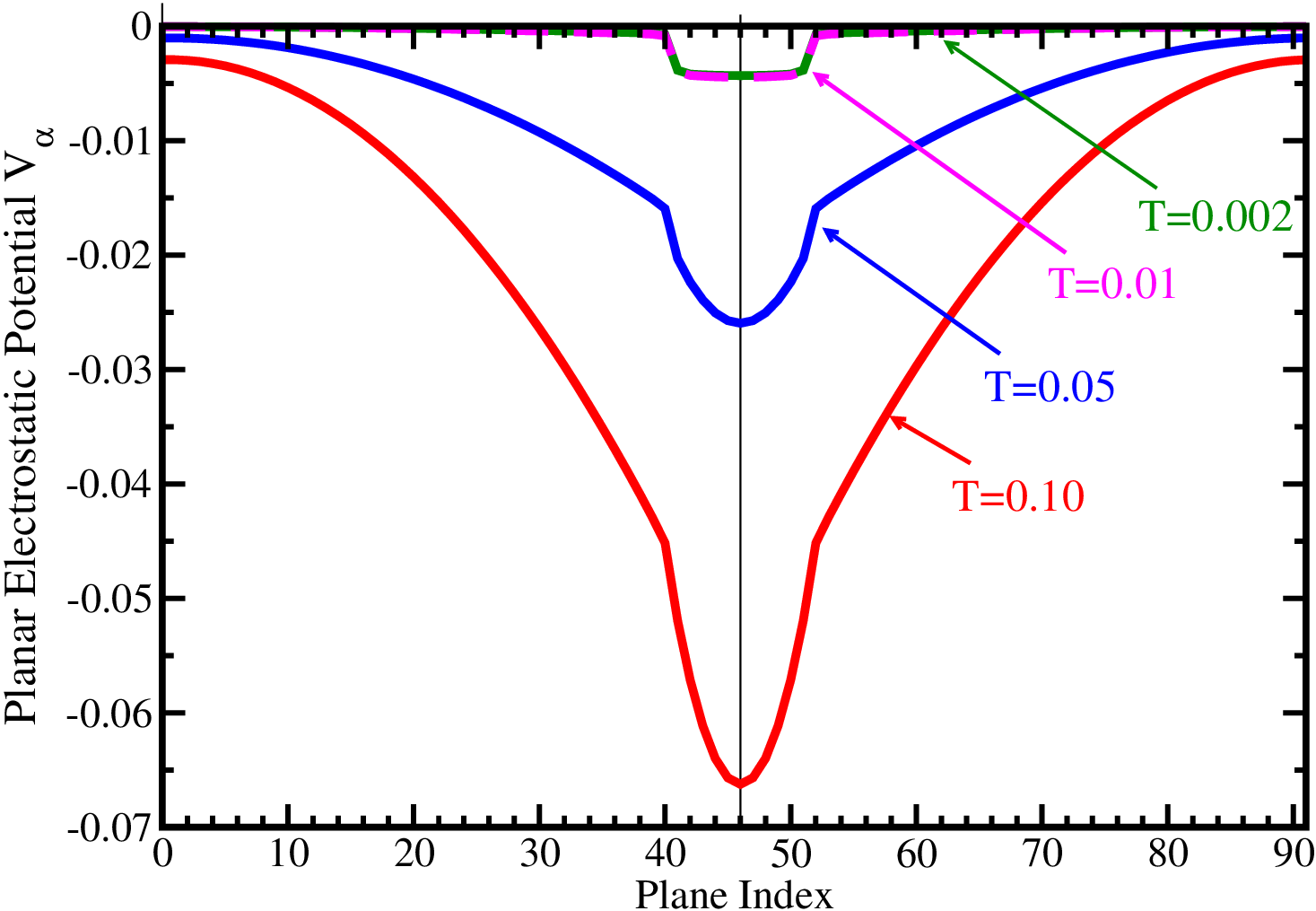}
\caption[]{(color online) 
The reconstructed charge, $\delta n_\alpha=n_\alpha-n_{BG}$  and the electrical potential   $V_\alpha$ 
are plotted at various temperatures as functions of the planar index 
for the band shift  $\Delta E_{c}=-0.9$ in the central planes. 
The full black line, dashed magenta, full blue, and red lines correspond to $T=0.002 ,  0.001, 0.05$, and $0.1$, respectively. 
Left panel: The reconstructed charge  $n_\alpha$.
Right panel: The potential $V_\alpha$ given by Poisson's equation. 
\label{fig:charge_and_potential_T=0.01} }
\end{figure*}

{\subsection{Electronic charge reconstruction and renormalization of the electrochemical potential}\label{charge reconstruction}}
The electronic charge distribution and  renormalized electrical potential are obtained by solving self-consistently the DMFT  and Poisson's equations 
on the imaginary axis. The results are illustrated in Fig.~\ref{fig:charge_and_potential_T=0.01}, 
where $\delta n_\alpha=n_\alpha-n_{BG}$ and $V_\alpha$ are plotted versus planar index $\alpha$ for several temperatures. 
For the planes in the leads, the parameters are  $U=8$, $\Delta E=0$ and $\epsilon_{Schott}=0.4$, while for the central planes we have $U=8$, $\Delta E=-0.9$, and $\epsilon_{Schott}=1.0$. 
The charge $n_\alpha$ deviates most strongly from the bulk values for the planes around the interface, where a screened-dipole layer forms, 
such that $ \delta n_{\alpha^+} \simeq -\delta n_{\alpha^-}$, and the curvature of the local potential changes sign. 
The potential due to the electronic charge reconstruction is large in the middle of the ML and small  on the first and the last self-consistent plane. 
The corresponding electric field decays away from the screened-dipol layer, with the decay length controlled by $\epsilon_{Schott}$: 
the larger the value of $\epsilon_{Schott}$, the smaller the decay length. 
The profile of the renormalized charge in the central part of the ML depends on the number of planes. 
By looking at the devices with M=5, 11, 21, and 31 central planes, we find that the charge on the central (mirror-symmetric) plane 
rapidly approaches $n_{BG}$,  as M increases. 

\begin{figure*}[thb!]
\centering
\includegraphics[width=0.3\columnwidth,clip]{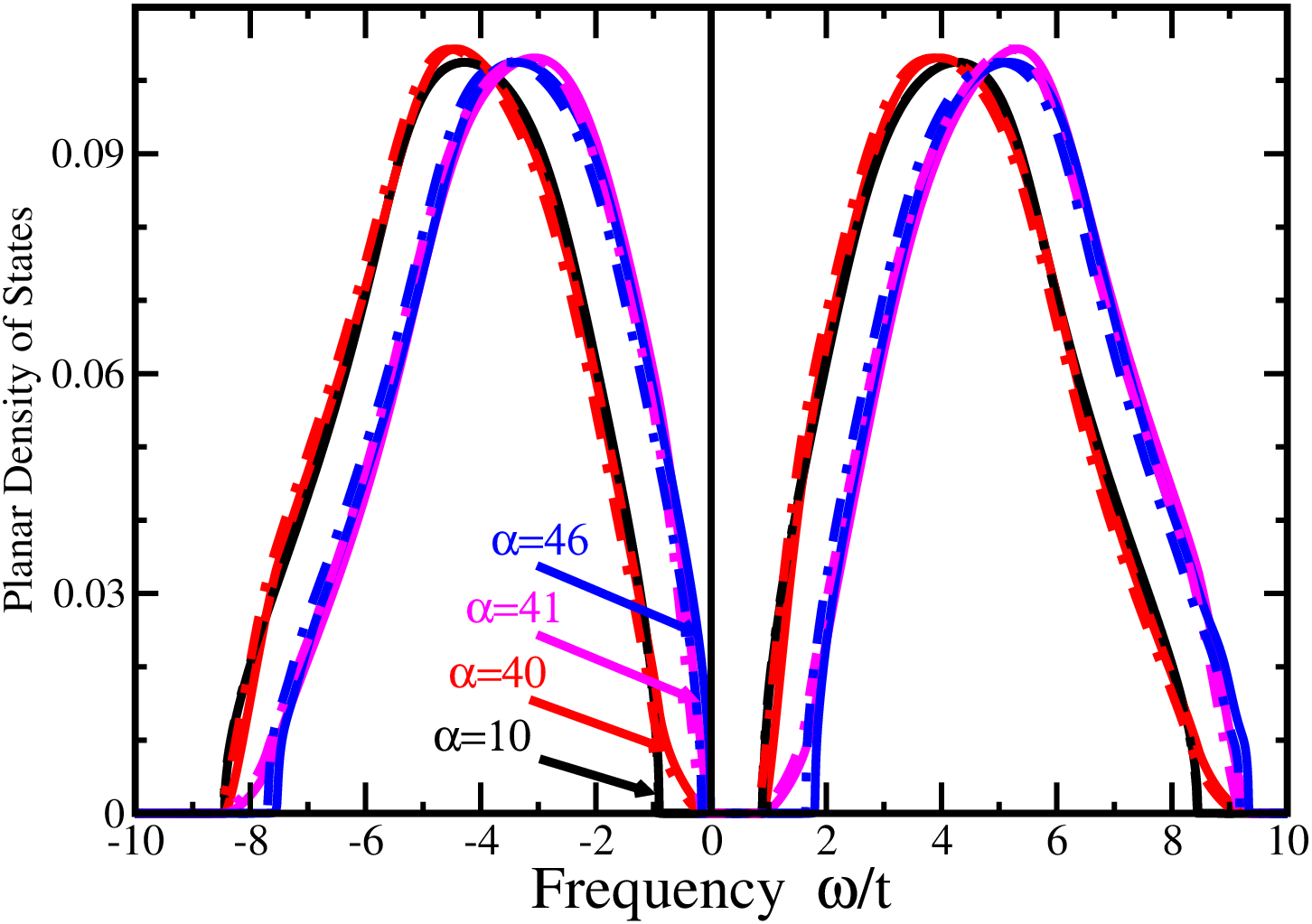}
\includegraphics[width=0.3\columnwidth,clip]{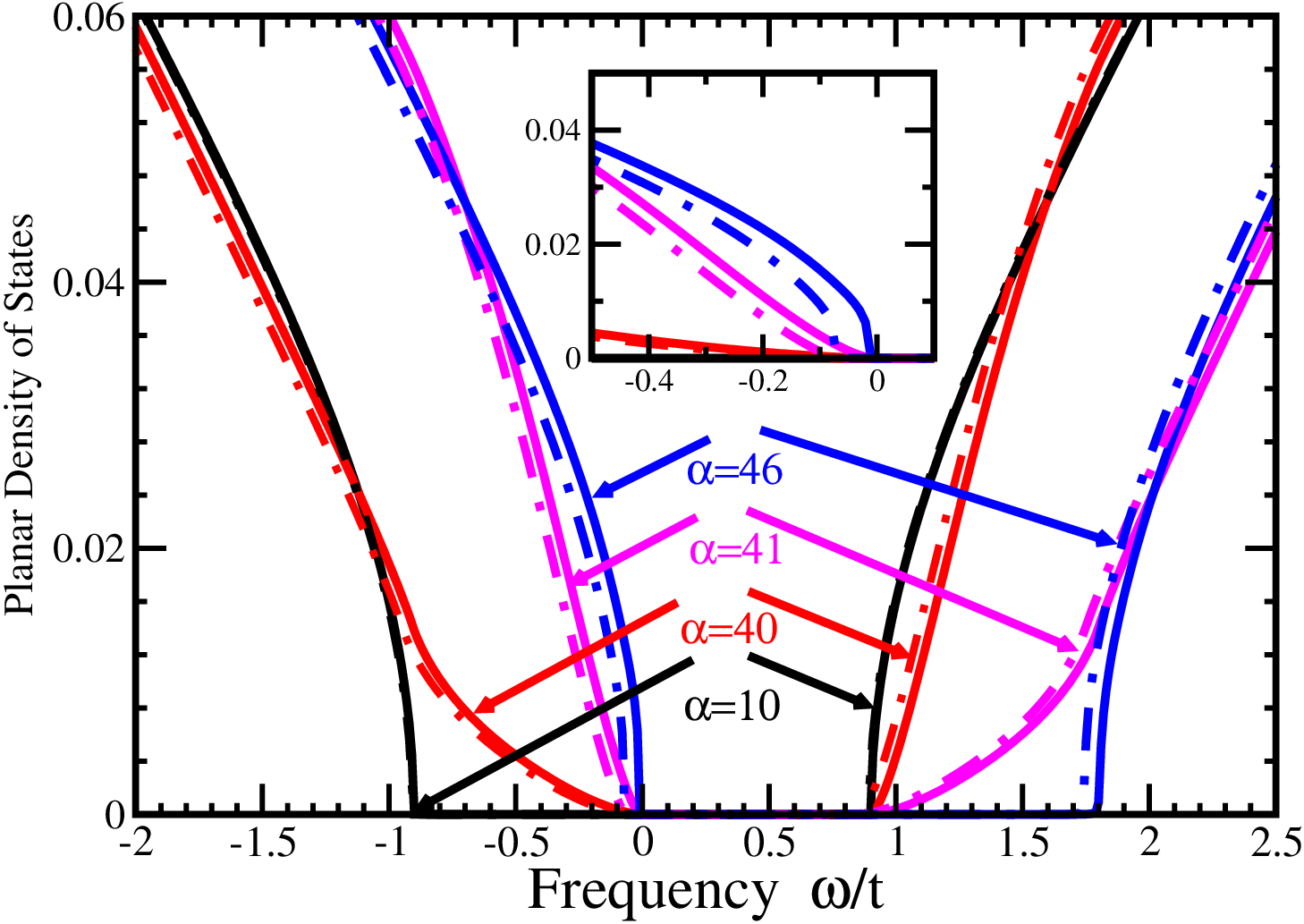}
\includegraphics[width=0.3\columnwidth,clip]{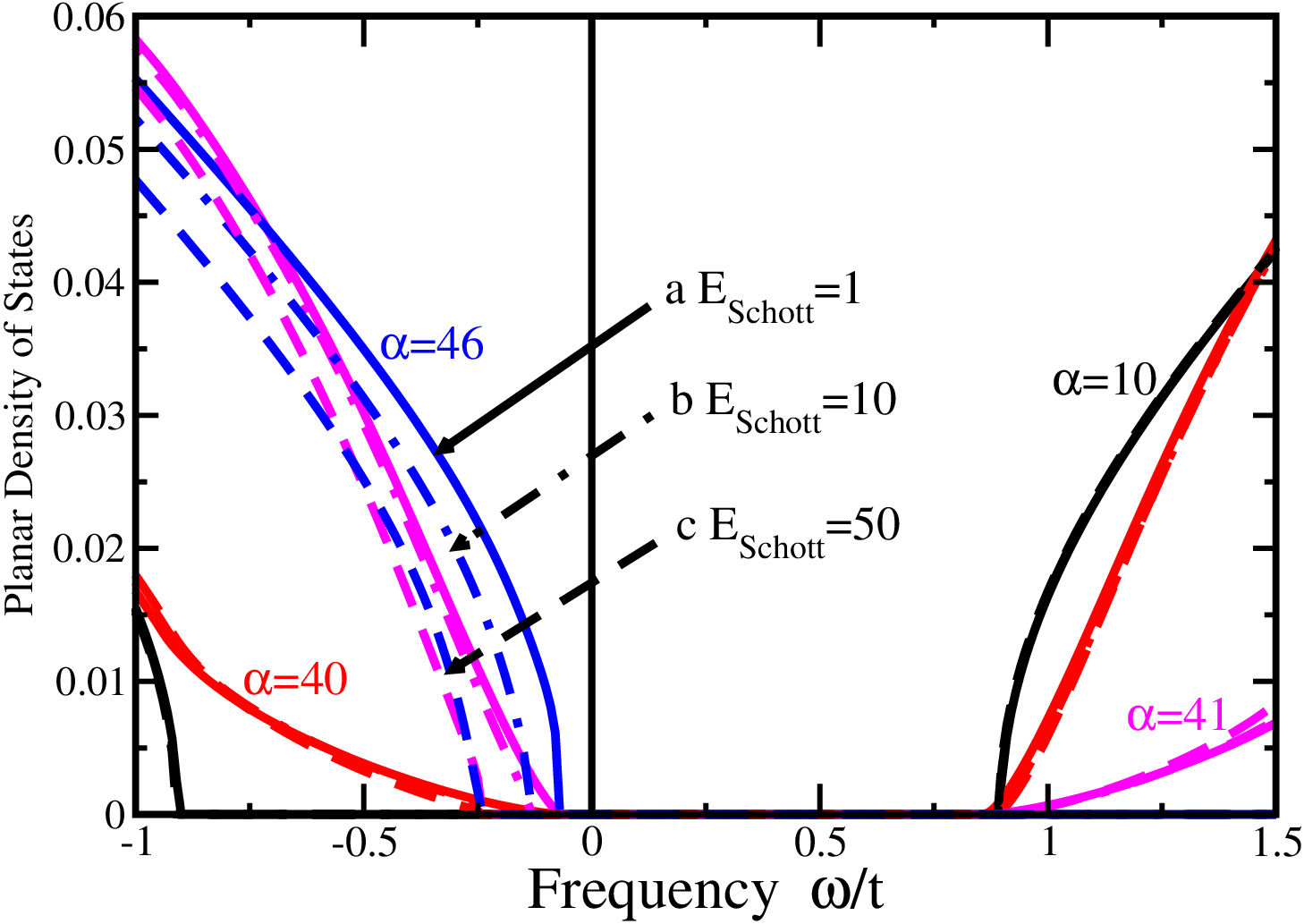}
\caption[]{(color online) 
Left panel:  
The local DOS, calculated at $k_BT=0.003$  (full lines) and $k_BT=0.1$  (dashed-dotted lines) 
is plotted, for $\Delta E_{c}=-0.9$ and constant $\epsilon_{Schott}$ in the leads and the center, as a function of frequency for several planes of the ML.  
The black lines show the plane in the lead which is far away from the interface ($\alpha=10$), 
red lines show the last plane in the left lead ($\alpha=40$), 
the magenta lines show the leftmost plane in the central part ($\alpha=41$), 
and the blue lines show the mirror plane of the ML ($\alpha=46$). 
Central panel:
The low-frequency part of the local DOS is shown for the same planes as in the left panel. 
The inset shows the very-low frequency part of the planar DOS. 
Right panel: 
$\rho_{\alpha}$ plotted versus $\omega$ for $k_BT=0.1$, $\Delta E_{c}=-0.9$, $\epsilon_{Schott}=0.4$ in the leads, and various values of $\epsilon_{Schott}$ in the center.   
The blue, magenta, red, and black curves show the local DOS on the mirror plane ($\alpha=46$), the plane next to the interface ($\alpha=41$ and $\alpha=40$, 
and the plane deep in the leads ($\alpha=10$), respectively. 
The values of $\epsilon_{Schott}$ in the center are $1.0$ (full lines), $10$ (dashed-dotted lines), and $50$ (dashed lines), respectively . 
}
\label{fig:dos_T=0.01}
\end{figure*}

Overall charge neutrality is well satisfied at low temperatures, where $\delta n=\sum_\alpha \delta n_\alpha\ll  \vert\delta n_{\alpha^+}\vert$. 
However, as  temperature increases, the error in the local charge distribution grows and the total charge does not sum to zero, as it must. 
We find that the charge accumulates on the central planes, as shown clearly in Fig.\ref{fig:charge_and_potential_T=0.01}. 
The screened-dipole layer is still well defined but, for $T= 0.1$, the charge accumulated in the central part becomes comparable 
to the charge forming the screened-dipole layer,  so that the renormalization of the electrical potential cannot be neglected at the edges of the ML. 
Even when the total accumulated  charge is small with respect to $n_\alpha$, the ensuing long-range potential shifts 
all the electron states of the ML, which inhibits the approach of our iterative procedure to the fixed point. 
We believe, the numerical instability arises because the chemical potential on the central planes is in the gap and very close to the band edge 
(see middle panel in Fig.~\ref{fig:dos_T=0.01},  Sec.~\ref{Local DOS}). Thus, even a very small shift of the electro-chemical potential, enforced by the 
Poisson's equation in the $n$-th iterative step, changes the occupancy of the central planes by too much, 
so that the $(n+1)-$th iterations cannot take us away from the incorrect solution with an overall net charge in the system. 
By adjusting the rate at which iterations are updated and using, at a given temperature, slightly different initial conditions,  
one can stabilize the system to some degree, but eventually it appears that the iterative method fails.
A large charging error is an indication of a breakdown of our iterative scheme and the data set $\{n_\alpha, V_\alpha\}$ with this behaviour 
is only a rough approximation to the actual solution.
If the band offset is reduced below $\vert \Delta E\vert < 0.9$, the  charge sum-rule holds to higher temperatures but the charge in the screened-dipole layer gets smaller. 
Eventually, for $\Delta E \to 0$,  
the system becomes uniform, and the screened-dipole layer vanishes. 

The behaviour displayed in Fig.~\ref{fig:charge_and_potential_T=0.01} can be related to the renormalization of the electron excitations, as  discussed in the next section. 
Here, we just remark that for the parameters used in that figure, the chemical potential is in the gap, 
for all the planes of the ML (see Fig.~\ref{fig:dos_T=0.01} in Sec.~\ref{Local DOS}).  
Denoting  by $\Gamma$ half the value of the Mott-Hubbard gap and by $\Delta$ the separation between the chemical potential and the nearest band edge, 
we find $\Delta\simeq \Gamma$  for the leads and $\Delta\ll \Gamma$ in the center. At low temperatures, the Fermi window is smaller than 
$\Delta$ and the system behaves as an insulator. For large enough temperatures, the Fermi window exceeds $\Delta$ and thermal fluctuations 
give rise to a  metallic behaviour which, however, is difficult to reach by our iterative procedure. 
This shifting of the band edges in the central region is the mechanism that enables the system to ``self-dope" 
and leads to many of the interesting transport phenomena that we present below.

\begin{figure*}[thb!]
\centering
\includegraphics[width=0.40\columnwidth,clip]{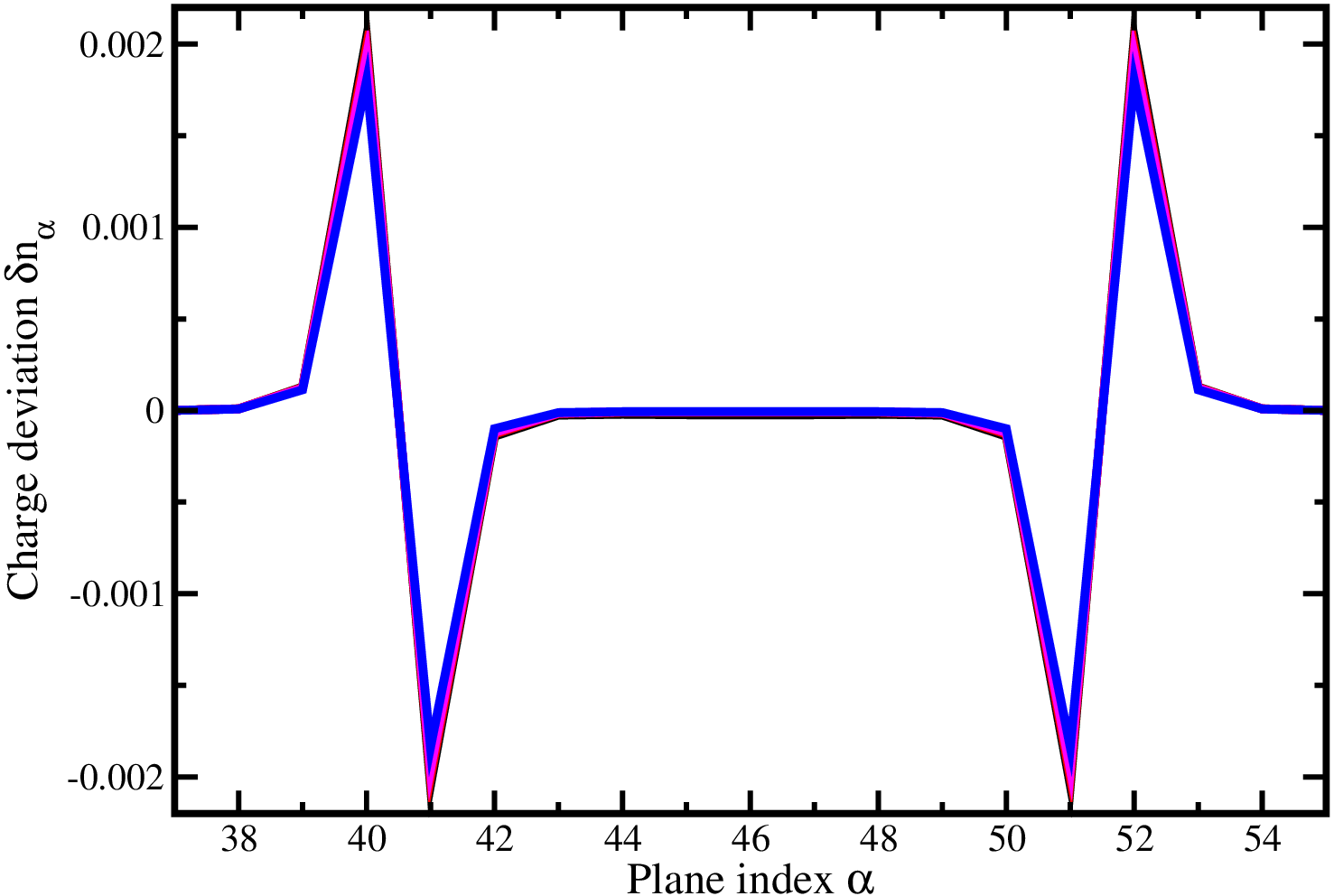}
\includegraphics[width=0.40\columnwidth,clip]{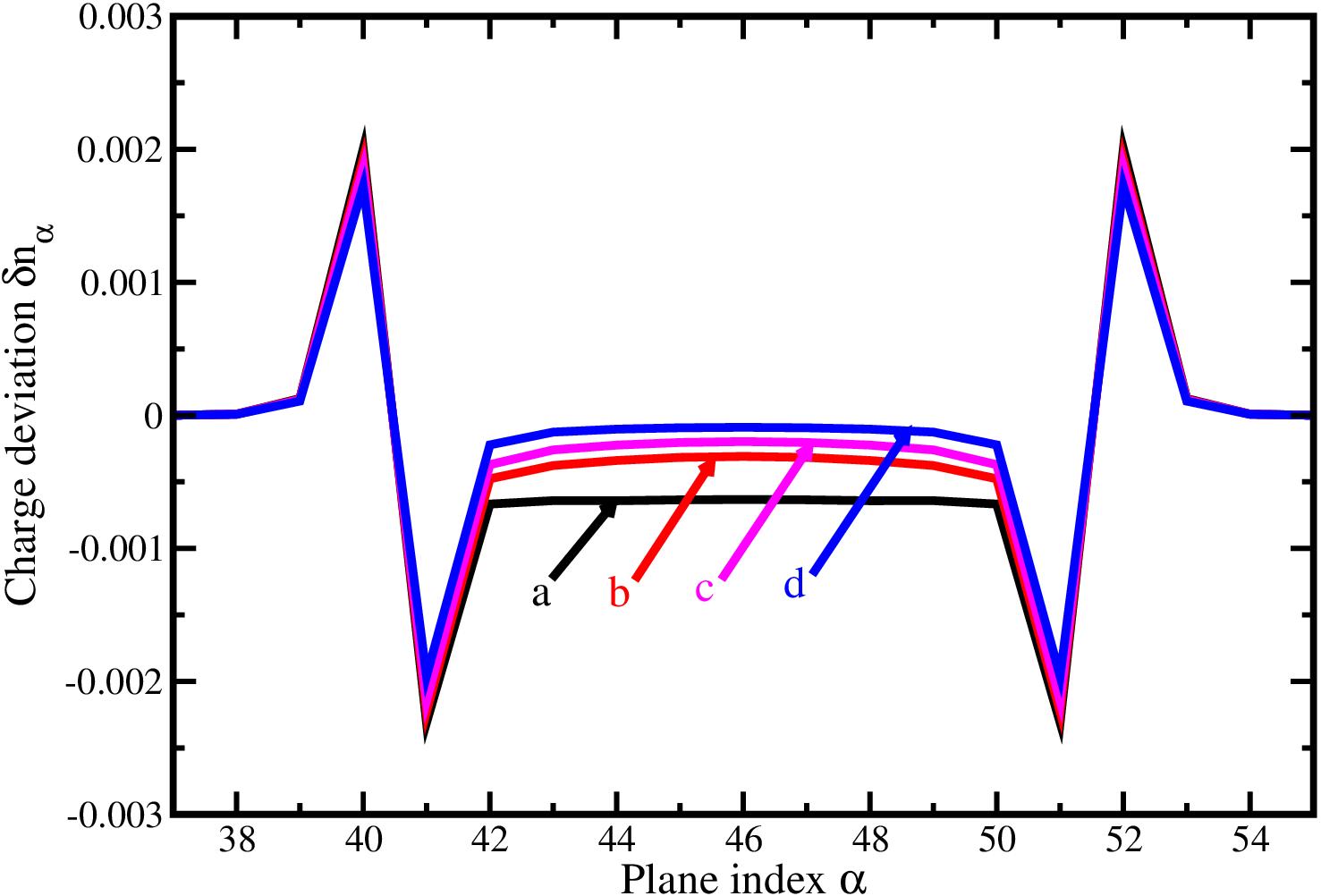}
\caption[]{(color online) 
The reconstructed charge, $\delta n_\alpha=n_\alpha-n_{BG}$ is plotted versus planar index $\alpha$ for various values of the screening parameter 
$\epsilon_{Schott}$ on the central planes. The other parameters are the same as in Fig.\ref{fig:charge_and_potential_T=0.01}.  
Left panel: $\delta n_\alpha$ for $T=0.01$.
Right panel: $\delta n_\alpha$ for  $T=0.1$; 
the lines a, b, c, d are computed with $\epsilon_{Schott}=1, 10, 20$, $50$, respectively. The corresponding charging errors ($\delta n \times 10^{3}$) 
are  -0.18  -0.09,  -0.08, and  -0.07 for $T=0.01$ and  -6.14, -3.66,  -2.61, and -1.41 for $T=0.1$. 
\label{fig:charge_eschott} }
\end{figure*}
The charging error is reduced and the solution extends to higher temperatures, if the screening parameter $\epsilon_{Schott}$ is increased in the central part. 
This is illustrated in Fig. \ref{fig:charge_eschott}, where $\delta n_\alpha$ is shown for $\Delta E_{c}=-0.9$ and various values of  $\epsilon_{Schott}$ in the center. 
At low temperature, where the charging error is small (see the left panel), an increase of $\epsilon_{Schott}$ doesn't affect the quality of the solution but 
at higher temperatures, where thermal fluctuations delocalize the conduction electrons and increase the metallicity of the central part, 
the charging error is systematically reduced as $\epsilon_{Schott}$ increases (see the right panel). 
For the data in Fig. \ref{fig:charge_eschott}, we find that the charging error at $T=0.01$ is much less than the charge in the screened-dipol layer, 
while at $T=0.1$ it is comparable or larger that $\delta n$. To reduce the charging error and obtain an acceptable solution, 
the screening length in the central part has to be reduced by a factor of 50 or more.  
In principle, the temperature dependence of the screening length should be taken into account when calculating the properties of the ML. 
In a fully self-consistent scheme, one  would not just renormalize the electron charge but would also renormalize the dielectric constant in Poisson's equation.  
One way of doing this would be to calculate the dielectric constant from the charge susceptibility which is known exactly for the Falicov-Kimball model\cite{freericks_2003}. 
However, any additional integrations make our current iterative scheme too slow, so that the renormalization of the dielectric constant requires 
a new numerical approach. In what follows, we restrict the calculations to constant screening parameters, $\epsilon_{Schott}=0.4$ in the leads and $\epsilon_{Schott}=1.0$ in the center 
and control the error by staying at relatively low temperatures.  
  
{\subsection{Local DOS}\label{Local DOS}}
The  frequency dependence of the local DOS and transport DOS is obtained, for a given set $\{n_\alpha, V_\alpha\}$,  
by  solving the DMFT equations on the real axis. 
The results obtained for the same parameters as in Fig.~\ref{fig:charge_and_potential_T=0.01}, 
are plotted  in the left and the central panel of Fig.~\ref{fig:dos_T=0.01}, for several typical planes of the ML. 
The full and dashed lines show the data obtained for $k_BT=0.003$ and $k_BT=0.1$, respectively. 
The left panel shows the overall behaviour of $\rho_{\alpha}(\omega)$  for $\alpha=10$ (black curve), $\alpha^-=40$ (red), 
$\alpha^+=41$ (magenta), and $\alpha_c=46$ (blue). 
These data are representative for the planes in the leads far away from the interface, the planes on the two sides of the interface, 
and the mirror plane of the ML. 
Because of the symmetry  with respect to the mirror plane $\alpha_c$, we have $\rho_{\alpha_c+\beta}(\omega)=\rho_{\alpha_c-\beta}(\omega)$ 
and do not have to show the data for $\alpha > \alpha_c$.
For $\alpha=10$, the local DOS is independent of temperature and it doesn't change with the band shift in the central part; 
$\rho_{10}(\omega)$  is a symmetric function of $\omega$, which is unaffected by the interface and it is almost the same as  $\rho_{bulk}(\omega)$. 
The same features are found for all other planes in the leads, except for the two planes next to the interface. 
Similarly, the local DOS on all the planes in the central  part of the ML is almost the same as  $\rho_{\alpha_c}$, 
except for the planes next to the interface.

The local DOS on the mirror plane and the local DOS on the leftmost plane in the leads are related 
by a simple shift, $\rho_{\alpha_c}(\omega)=\rho_0(\omega+a(T)\Delta E_{})$, 
where $a(T)$ is weakly temperature dependent coefficient of the order of one. 
The same relationship holds for all other planes in the center, except for the planes next to the interface which are not just shifted but distorted as well. 
Thus, while the local DOS on the planes in semi-infinite leads are nearly the same as in the bulk, 
the excitations in the central part of the ML are pushed almost rigidly to higher energies. 
As regards the planes close to the interface, their local DOS is asymmetrically distorted with respect to the bulk, 
as shown  in the central panel of Fig.~\ref{fig:dos_T=0.01}. 
The distortion of the occupied part of   $\rho_{\alpha^-}(\omega)$ is consistent with the increased number 
of local electrons on the plane $\alpha^-=40$ and the distortion of the unoccupied part of  $\rho_{\alpha^+}(\omega)$ 
is consistent with the reduced number of local electrons on plane $\alpha^+=40$. 

The interfacing has a pronounced effect on the excitation spectrum of the ML. 
Unlike in the bulk, where the chemical potential is completely determined by the number of electrons,  
so that the local DOS of a half-filled system is symmetric with respect to the chemical potential,  
in the heterogeneous ML, the energy levels of electrons on the central planes are shifted with respect to the chemical potential, 
even though the number of electrons on most planes is the same as in the bulk. 
This is not surprising, since the electrons on the central planes have to adjust to the chemical potential set by the leads,  
while trying to maintain the local charge neutrality and minimize the potential energy. 
The shift of the renormalized excitation energies on the planes in the central part depends to some extent on the electrostatic potential $V_\alpha$ 
and the real part of the self-energy but the main contribution is coming from the shift of the local chemical potential by $\Delta E$.  
By tuning the offset $\Delta E$ we can bring the chemical potential arbitrarily close to the band edge and still keep 
most of the planes at half-filling; the charge neutrality is only broken at the planes that form an electrical dipole at the interface 
and give rise to the electrical potential that has a minimum (or maximum) in the center of the ML and vanishes at the edges.
 
For $\Delta E <0$, the energy of the lowest unoccupied single-particle state is fixed by the leads 
(see the black curves in Fig.~\ref{fig:charge_and_potential_T=0.01}), 
while the energy of the highest occupied state is fixed by the central planes (see the blue curves in Fig.~\ref{fig:charge_and_potential_T=0.01}). 
Since the chemical potential  (zero of energy) is also fixed by the  leads, the upward shift of the local DOS on the central planes 
reduces the overall magnitude of the gap in the excitation spectrum. 
Furthermore, even though the chemical potential is in the gap, the separation $\Delta$ between 
the chemical potential and the band edge of the lower Hubbard band is greatly reduced 
with respect to the bulk (see the black and blue curves in left and central panels in Fig.~\ref{fig:dos_T=0.01}). 
While temperature doesn't affect much the local DOS in the leads, an increase of temperature shifts $\rho_{\alpha_c}(\omega)$ 
to lower frequencies,  and shifts and distorts $\rho_{\alpha^-}(\omega)$ and $\rho_{\alpha^+}(\omega)$. 
The band edge of the central planes moves away from the chemical potential and we find that $\Delta$ increases with temperature  in sub-linear fashion.  
This temperature-dependent shift of the local DOS is caused by the temperature dependence of the Fermi factors {\em and} 
the renormalization of the electrochemical potential. 
As shown by the inset in Fig.~\ref{fig:dos_T=0.01},  we have $\Delta > k_B T$  for $T=0.03$ and  $\Delta < k_BT$ for $T=0.1$. 

The same features are observed for other choices of the parameters. 
For $\Delta E > 0$,  the energy of the highest occupied single-particle state is fixed by the leads, while the energy of the lowest 
unoccupied state is fixed by the central planes, so an increase of $\Delta E_{}$  reduces the gap and $\Delta$. 
A large enough $\vert\Delta E\vert$ brings the chemical potential into the lower or the upper Hubbard band and makes the central planes metallic. 
Thus, for an appropriate choice of model parameters, a small gate voltage applied to the central part can switch the system from a metallic to an insulating state. 
Unfortunately, our current iterative solution of the DMFT equations cannot provide the details of the transition. 

If we take constant screening parameters and perform the calculations at elevated temperature for a large band offset, 
say, $T=0.1$ and $\Delta E =-0.95$,  the chemical potential is in the conduction band and the central part is metallic.  
A reduction of temperature shifts the band edge across the chemical potential and gives rise eventually to a metal-insulator (MI) transition. 
However, for $k_B T\geq \Delta$, the charging error is large and  the renormalized  electrical potential doesn't vanish at the 
edge of the sample (see Fig.\ref{fig:charge_and_potential_T=0.01}). 
Close to the MI transition, which is the most interesting case for applications, we can evaluate the precise temperature dependence 
of the local DOS, as long as $k_B T\leq \Delta$. 
The high-temperature DOS can only be discussed in a qualitative way, because it is obtained by the real-axis calculations which 
rely on the values of $n_\alpha$ and $V_\alpha$. These data are produced by the imaginary-axis calculations which have, for $k_B T\geq \Delta$,
a large error bar. However, by studying a few cases where the iterative solution produced a small charging error, 
we could see that the real-axis solution  exhibits qualitatively the same features as above. In other words, the real-axis results are robust with 
respect to the small errors in the imaginary-axis data. 

As mentioned in the previous section, the error in the charge distribution can be greatly reduced by increasing 
$\epsilon_{Schott}$ in the central part (see  Fig.~\ref{fig:charge_eschott}). 
The effect of $\epsilon_{Schott}$ on the local DOS is shown in  the right panel of Fig.~\ref{fig:dos_T=0.01}, 
where $\rho_{\alpha}(\omega)$ is plotted for $T=0.1$, $\Delta E =-0.9$, $\epsilon_{Schott}=0.4$ in the leads, and 
$\epsilon_{Schott}=1$, 10, and 50  in the central part.  The data show that an increase of $\epsilon_{Schott}$, at fixed temperature,  
brings us from the regime where $k_B T \geq \Delta$ and the charging error is large (curve (a), $\epsilon_{Schott}=1$)  
to the regime  where $k_B T \leq \Delta$ and the charging error is small (curve (c), $\epsilon_{Schott}=50$). 
This feature can be used to extend the temperature range in which the transport coefficients can be computed.

The reduction of the Coulomb interaction in the leads makes the gap on the A-planes smaller than on the B-planes and further reduces 
the separation between the lower band edge (set by the B-planes) and the upper band edge (set by the A-planes).  
By tuning the parameters, we can transform a Mott insulator into a narrow gap semiconductor with interesting thermoelectric 
properties. The deviation of the electron number from 1/2, found for $\alpha^-$ and $\alpha^+$, the two planes next to the interface, 
is accompanied by the deviation of $\rho_{\alpha^-}(\omega)$ and $\rho_{\alpha^+}(\omega)$ from the bulk shape, which further affects
the thermoelectric response. 
\\

\subsection{Transport function and transport coefficients}
\begin{figure*}[thb!]
\centering   
\includegraphics[width=0.4\columnwidth,clip]{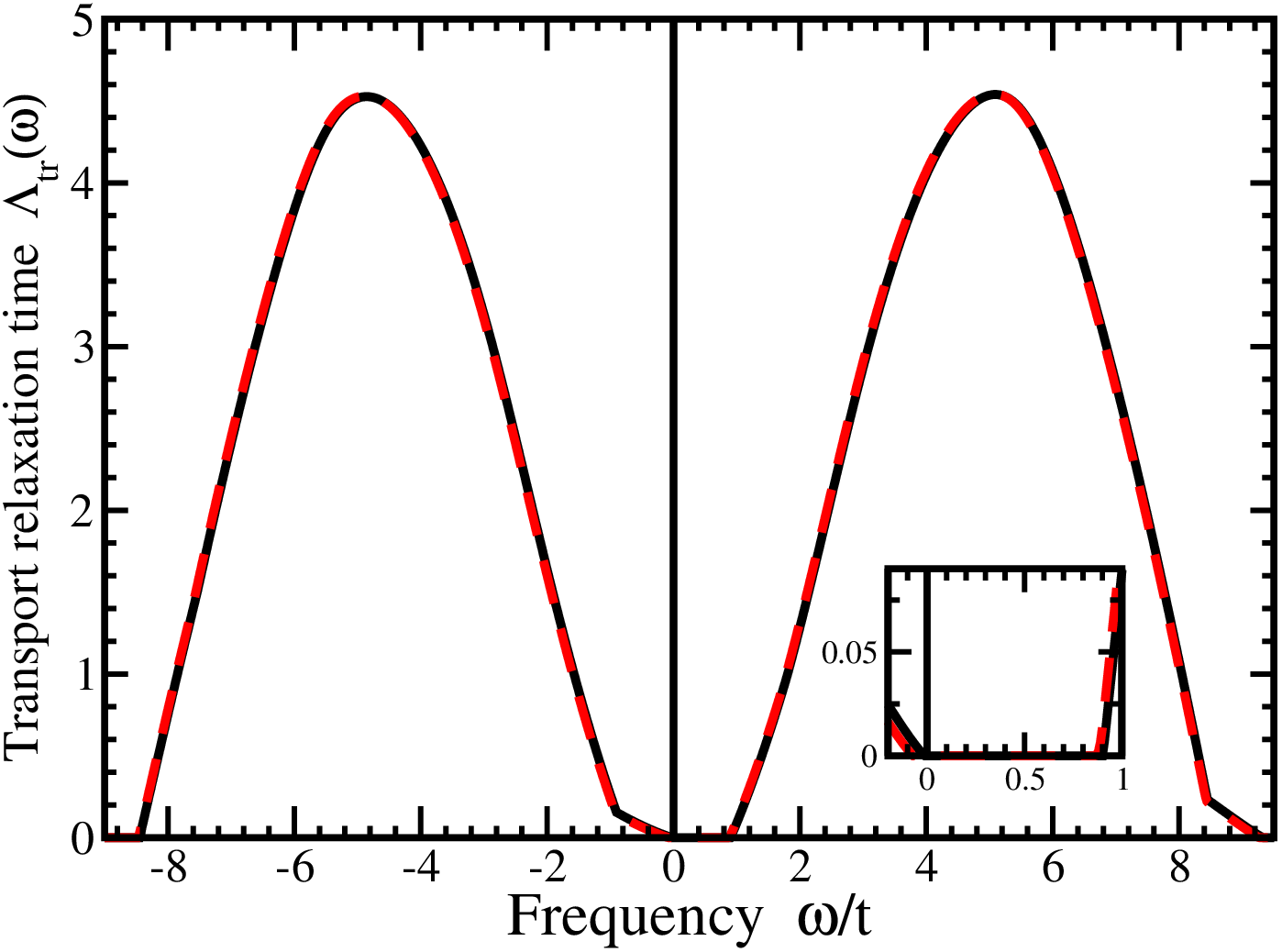}
\caption[]{(color online) 
The transport function calculated at $k_BT=0.003$ (full black line) and $k_BT=0.1$ (dashed red line) 
is plotted, for $\Delta E_{c}=-0.9$, as a function of frequency.  
Inset:  The low-frequency part of $\Lambda_{tr}(\omega)$ showing the shift of the transport function to higher energies 
and the reduction of the transport gap at low temperatures. 
}
\label{fig:transport_function}
\end{figure*}

The overall features of the transport function of the ML, calculated for $U=8$, $\epsilon_{Schott}=0.4$ and $\Delta E=0$ in the leads,  
and $U=8$, $\epsilon_{Schott}=1.0$ and $\Delta E=-0.90$   in the center, are shown in Fig.~\ref{fig:transport_function}, 
where $\Lambda_{tr}(\omega)$ is plotted versus frequency for $k_BT=0.003$ and $k_BT=0.1$. 
The transport gap is greatly reduced by the charge reconstruction and the ensuing long range potential 
which gives rise to an upward shift of the electron excitations on the central planes. 
In the bulk,  the transport density of states of the Falicov-Kimball model is temperature-independent, when the concentration of $f$-electrons is constant, while 
in a heterogeneous multilayer, we find that $\Lambda_{tr}(\omega)$ is weakly temperature-dependent. 
The low-energy part of  $\Lambda_{tr}$, relevant for the transport properties, is shown in the inset of Fig.~\ref{fig:transport_function}. 
The slight temperature-induced shift of $\Lambda_{tr}$, partly due to the charging error, affects the transport coefficients in a quantitative but not in a qualitative way. 
Unlike in heavy fermions or other strongly correlated systems with a Fermi liquid ground state, here, 
the slope of the transport  function in the vicinity of the chemical potential is large. 
This has a drastic effect on the transport properties of the ML, as we now discuss.

\begin{figure*}[thb!]
\centering
\includegraphics[width=0.40\columnwidth,clip]{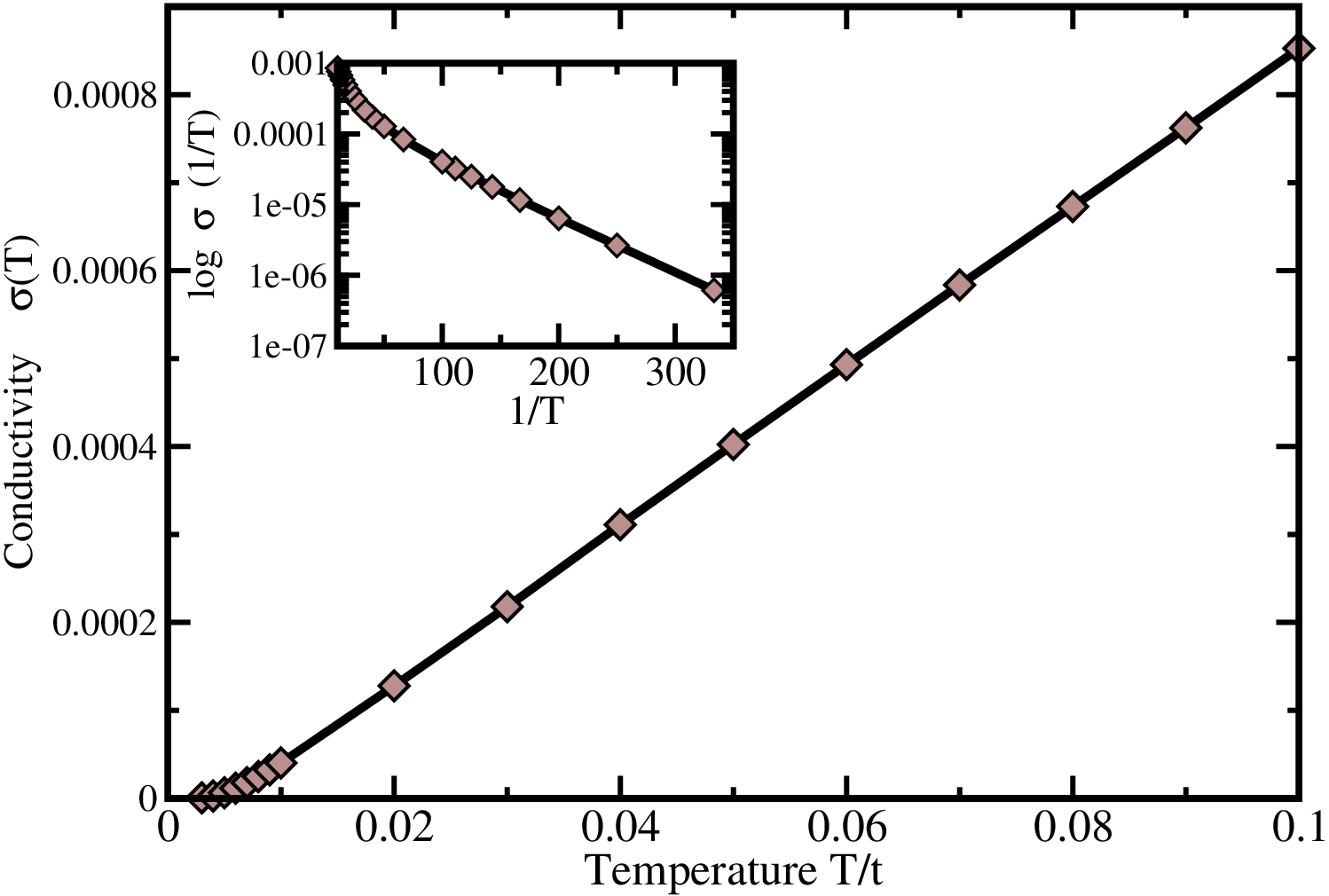}
\includegraphics[width=0.40\columnwidth,clip]{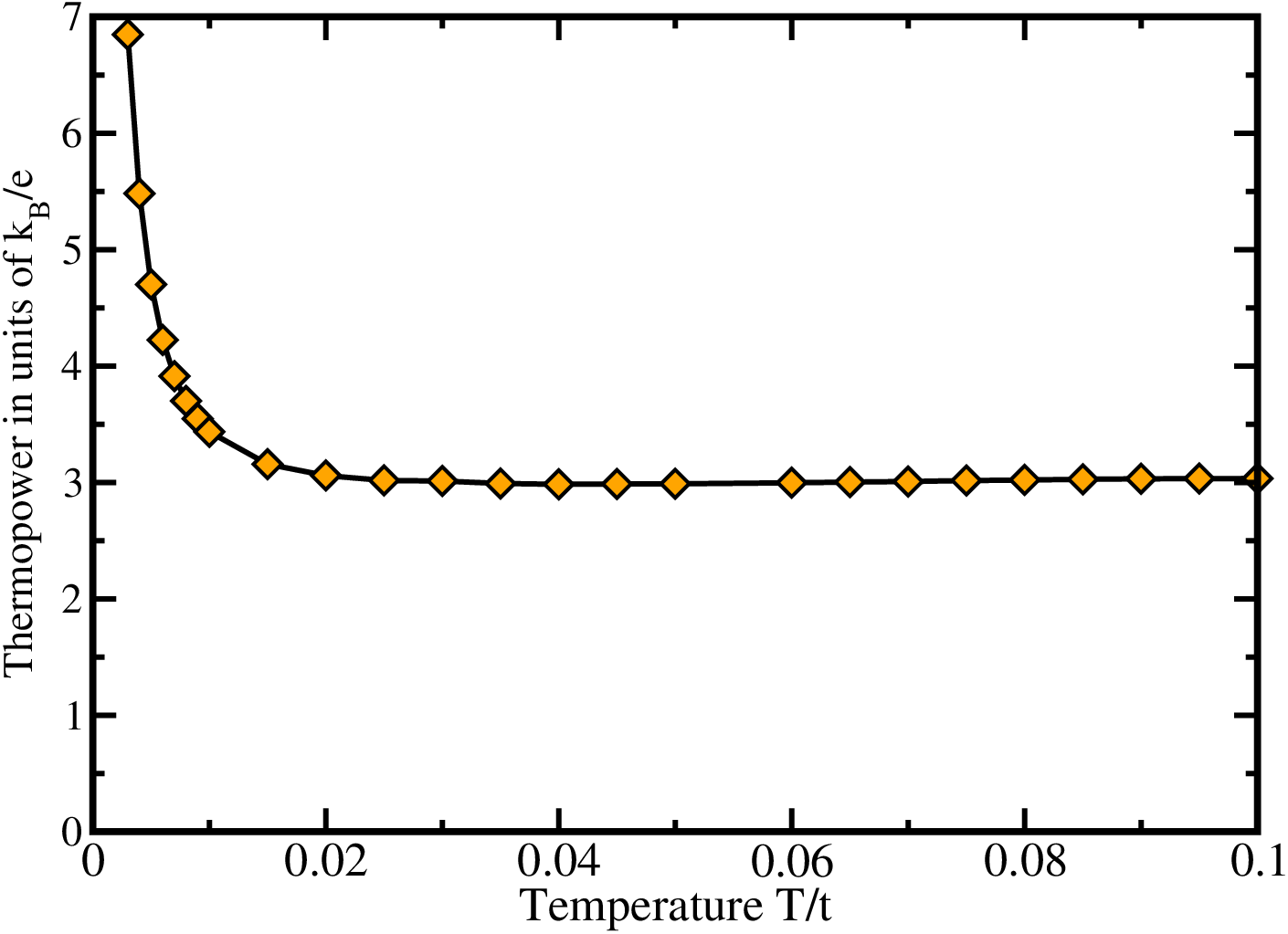}
\caption[]{Left panel: The conductivity of the ML is plotted as a function of temperature for the same parameters as in Fig.~\ref{fig:transport_function}.
Inset: the conductivity plotted on a logarithmic scale as a function of inverse temperature. 
Right panel: The thermopower of the ML in units of $[k_B/e]$ is plotted as a function of temperature for the same parameters as the conductivity.
}
\label{fig:conductivity} 
\end{figure*}
The temperature dependence of the electrical conductivity, computed for the same parameters as  
in Fig.~\ref{fig:transport_function}, is plotted in the left panel of Fig.~\ref{fig:conductivity}. 
At low temperatures, the conductivity exhibits an exponential decrease, $\sigma\simeq \exp(-1/T)$, as shown in the inset. 
By reducing the Coulomb repulsion on the central planes (and modifying the band offset, so as to keep the chemical potential 
close to the band edge), the conductivity can be further enhanced.
At high temperatures, $\sigma(T)$ shows a linear behaviour which is also found in homogeneous lightly doped Mott-Hubbard insulators, 
when the chemical potential is in the immediate vicinity of the band edge\cite{zlatic_2012,deng_2013,zlatic_2014}. 
However, the quantitative analysis of the inhomogeneous Mott-Hubbard materials is difficult because of the numerical charging error.  

The thermopower calculated for the same parameters as in Fig.~\ref{fig:transport_function} is shown in the 
right panel of Fig.~\ref{fig:conductivity}. 
At low temperatures, where the conductivity decreases, the thermopower increases rapidly. 
Close to the metal-insulator transition, the initial slope of the transport DOS at the chemical potential becomes very large 
and the thermopower of the multilayer can diverge. 
In the temperature range in which the electrical  conductivity is linear, the thermopower is nearly constant and still very large. 

\begin{figure*}[thb!]
\centering
\includegraphics[width=0.40\columnwidth,clip]{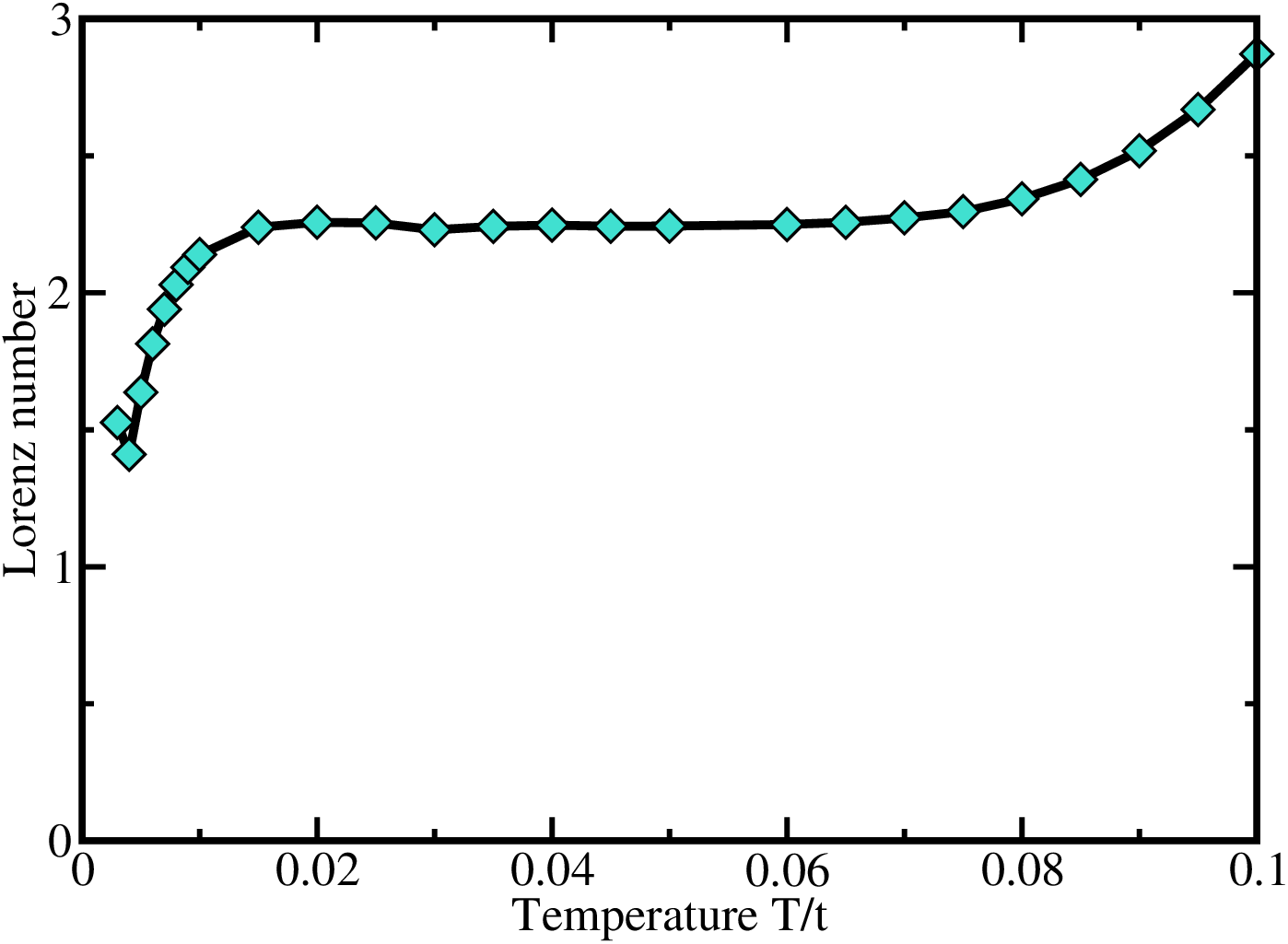}
\includegraphics[width=0.40\columnwidth,clip]{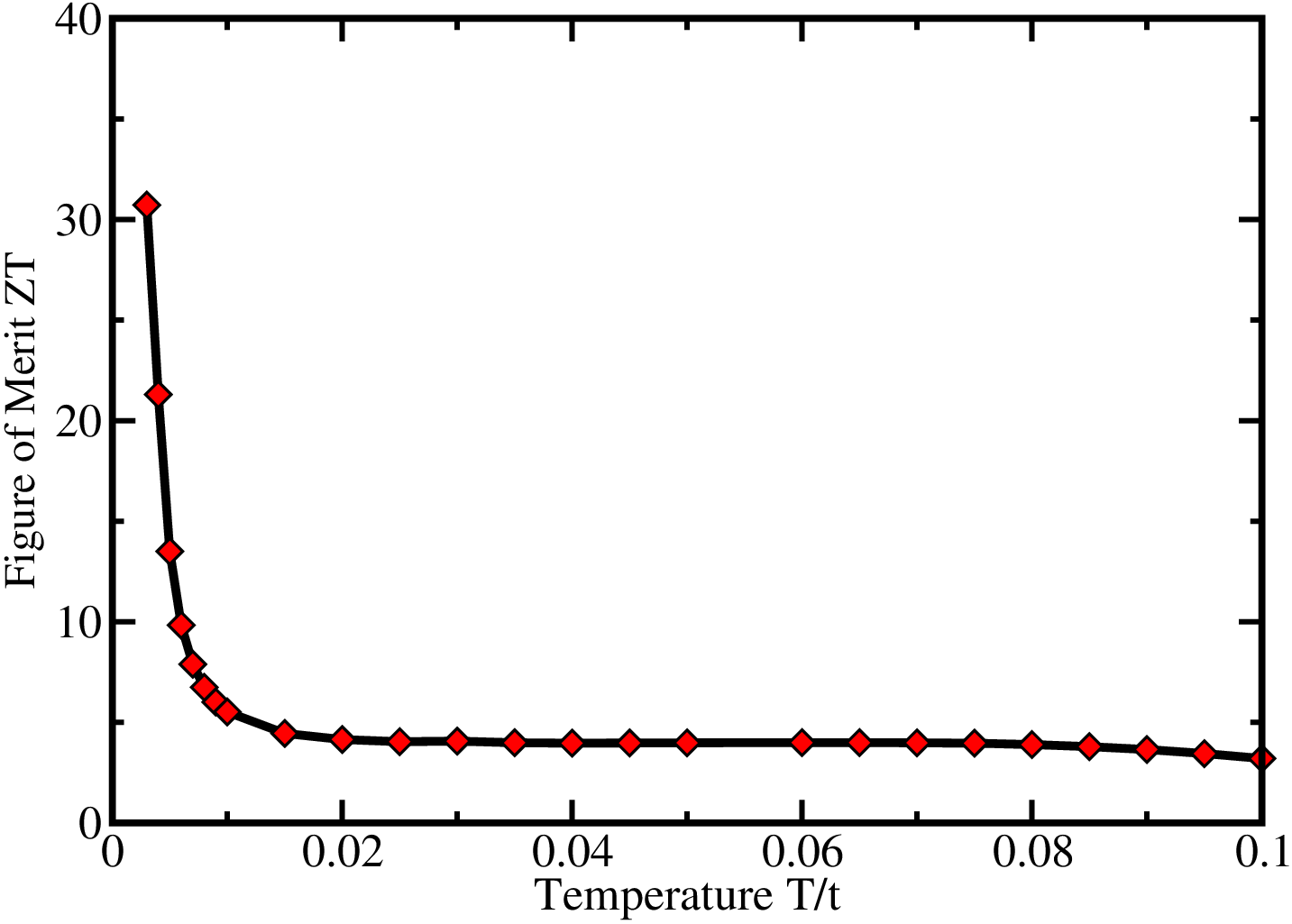}
\caption[]{
Left panel:  Lorenz number in units of $[k_B/e]^2$ plotted as  a function of temperature.
Right panel: ZT plotted as  a function of temperature. Parameters are the same as in Fig.~\ref{fig:conductivity}
}
\label{fig:zt} 
\end{figure*}

The effective Lorenz number plotted in the left panel of Fig.~\ref{fig:zt} shows at low temperatures large deviations 
from the universal value ${\cal L}_0=(\pi k_B)^2/(3e^2)$. 
Using the same parameters as in Fig.~\ref{fig:transport_function}, and neglecting the phonon contribution, 
we calculated  the figure-of-merit of the device, $ZT=\alpha^2/{\cal L}$. 
The results are shown in the right panel of Fig.~\ref{fig:zt}. 
In the temperature range where the thermopower becomes large and the Lorenz number 
deviates from its universal value, we find a large enhancement of the electronic figure-of-merit. 
When the device is close to the metal-insulator transition, a surprisingly large Seebeck coefficient 
and much enhanced figure-of-merit are observed. The enhancement is caused by the  
electron correlations and similar effects are not found for non-interacting electrons. 
The promising feature of the ML devices is the presence of the interface that can impede the phonon 
transport and improve the overall performance with respect to the bulk. 
If phonon scattering  is reduced due to the mismatch of ionic masses in the leads and the central planes, 
the thermoelectric performance of correlated multilayers could be improved by optimizing the number of planes in the central part 
so as to reduce the thermal conductivity. Unlike in bulk materials, where a large power factor (given by the product of the conductivity 
and the square of the thermopower) is usually accompanied by large thermal conductivity, in a ML the power factor and the thermal 
conductivity are caused by different processes and can be tuned independently. 

{\section{Summary and outlook}\label{summary}}
This paper presents a theory of charge and heat transport parallel to the interfaces of a ABA multilayer,   
where A and B are half-filled Mott-Hubard insulators with large gaps in their excitation spectrum.  
When separated, the chemical potentials of  A and B are located in the middle of their respective Mott-Hubbard gaps, 
while the energy bands of B are shifted relative to that of A by $\Delta E$. 
In a ML, the B material is sandwiched between A leads, so as to form an ABA structure. The leftmost and the rightmost A planes 
are attached to a semi-infinite bulk which sets the chemical potential of the device.
The excitations in the A leads are virtually unchanged by interfacing  but in the central part B, they are shifted by about $-\Delta E$ 
with respect to their position in the bulk.  
This gives rise to a charge redistribution and breaks the charge neutrality of the planes close to the interface. 
The ensuing electrical field couples self-consistently to the itinerant electrons, so that the properties of the ML crucially depend on an interplay 
between the on-site Coulomb forces and the long range electrostatic forces. 

We model  the ML by the Falicov-Kimball Hamiltonian and compute the Green's function and the local charge on each plane by inhomogeneous DMFT.  
The electrical potential corresponding to that charge is calculated from Poisson's equation and, in self-consistent calculations, 
it should coincide with the potential that determines the quantum states of conduction electrons.  
To ensure  the self-consistency we recalculate the Green's function and the charge distribution using the potential provided by Poisson's equation 
and iterate this procedure to the fixed point. 
On the imaginary axis, this procedure yields the equilibrium charge distribution and electrostatic potentials on each plane. 
Once they are determined, the real-axis calculations provide the excitation spectrum and the transport function of the heterostructure. 

In the leads, the electronic charge distribution and the planar DOS are almost the same as in material A, except for a few planes next to the interface, 
where, for $\Delta E<0$,  there is a charge accumulation. The local DOS on these planes deviates from the symmetric shape found in the bulk. 
In the central part, the local charge is also nearly the same as in material B, with the exception of the planes next to the interface, 
where the charge is reduced. The local DOS of the central planes retains its bulk shape but it is shifted almost rigidly  by  $-\Delta E$; 
the exceptions are the planes with a reduced charge, where the local DOS is not just shifted  but it is also distorted.  
The charge redistribution is mainly confined to the planes next to the A/B interfaces, where it gives rise to screened-dipole layers.   
Since the chemical potential and the energy of the lowest unoccupied states of the ML are set by the leads, 
while the energy of the highest occupied states is fixed by the central planes, and shifted by -$\Delta E$,  
the interfacing reduces the gap in the transport density of states and puts the highest occupied states just below the chemical potential. 
In such a heterostructure, neither charge nor heat  is transported perpendicular to the interface but there is a finite thermoelectric response parallel to it. 
This response, confined to the central part of the ML,  depends sensitively on the interplay between the on-site Coulomb repulsion and 
the long-range electrostatic forces. 

For the right choice of parameters, we find that a heterostructure built of two Mott-Hubbard insulators exhibits, in a large temperature interval, 
a linear conductivity and a large temperature-independent thermopower. Furthermore, we conjecture that the application of a temperature gradient 
perpendicular to the interface and a magnetic field parallel to the interface, say along the $x$-axis, would give rise to the Ettinghausen voltage or 
the Rigi-Leduc temperature drop (thermal Hall effect) along the $y$-axis. The experimental verification of these thermoelectric and 
thermomagnetic effects shouldn't be too difficult. 

Our results indicate that correlated ML's might be quite interesting for applications. First, by tuning the band-offset, so as to bring the gap edge in 
the immediate vicinity of the chemical potential, we can produce a heterostructure in which  a MI transition is easily induced by a small gate voltage.
The gate voltage does not perturb much the leads, where the chemical potential is in the centre of a large gap, 
but can render the central planes metallic. This switching does not involve the diffusion of electrons over macroscopic distances 
and it is much faster than in ordinary semiconductors.  Second, by selecting the parameters of materials A and B and tuning the number of 
planes in the central part, we can reduce the phonon scattering and thermal conductivity parallel to the planes without reducing the electrical 
conductivity or the thermopower. Thus, by combining strongly correlated materials with large electronic power factors but small ZT we can produce, 
theoretically at least, a heterostructure with a large ZT.  \\

\begin{acknowledgements}
We acknowledge useful discussions and critical remarks of R. Monnier.
This work is supported by the NSF grant No. EFRI-1433307. J.K.F. is also supported by the McDevitt bequest
at Georgetown University. V.Z acknowledges the support by the Ministry of Science of Croatia under the bilateral agreement 
with the USA on the scientific and technological cooperation, Project No. 1/2014. 
\end{acknowledgements}

\end{document}